\documentclass[onecolumn]{aastex}
\usepackage{emulateapj5}
\usepackage{epsf}

\slugcomment{Accepted for publication in The Astrophysical Journal, Vol. 562 p. 1,
2001 November 20}

\shorttitle{Cosmic Ray Electrons in GC}
\shortauthors{Miniati {\it et al.~}}

\def\cf{{\it cf.~}}
\def\eg{{\it e.g.,~}}
\def\ie{{\it i.e.,~}}
\def\lsim{\raise0.3ex\hbox{$<$}\kern-0.75em{\lower0.65ex\hbox{$\sim$}}} 
\def\gsim{\raise0.3ex\hbox{$>$}\kern-0.75em{\lower0.65ex\hbox{$\sim$}}}

\def\cm3{~{\rm cm^{-3}}}

\def\hinv{$h^{-1}$}
\def\phat{\hat{p}}

\def\cf{{cf.~}}
\def\eg{{e.g.,~}}
\def\ie{{i.e.,~}}
\def\lsim{\raise0.3ex\hbox{$<$}\kern-0.75em{\lower0.65ex\hbox{$\sim$}}} 
\def\gsim{\raise0.3ex\hbox{$>$}\kern-0.75em{\lower0.65ex\hbox{$\sim$}}} 
\def\sc1{\raise2.1ex\hbox{\tiny $r\!\!=\!\!4$}\kern-0.95em{\hbox{$=$}}}

\def\cm3{~{\rm cm^{-3}}}

\def\hinv{$h^{-1}$}

\def\phat{\hat{p}}

\def\ltsima{$\; \buildrel < \over \sim \;$}
\def\simlt{\lower.5ex\hbox{\ltsima}}
\def\gtsima{$\; \buildrel > \over \sim \;$}
\def\simgt{\lower.5ex\hbox{\gtsima}}

\def\sc{{\rm Science\ }}

\begin{document}

\title{Cosmic Ray Electrons in Groups and Clusters of Galaxies:
Primary and Secondary Populations from a Numerical Cosmological 
Simulation.\altaffilmark{6}}

\author{  Francesco Miniati     \altaffilmark{1,2,3},
          T. W. Jones           \altaffilmark{2},
          Hyesung Kang          \altaffilmark{4}, 
     and
          Dongsu Ryu            \altaffilmark{5}}

\altaffiltext{1}{Max-Planck-Institut f{\"u}r Astro\-phy\-sik,
Karl-Schwarz\-schild\--Str. 1, D-85741 Gar\-ching, Germany}
\altaffiltext{2}{School of Physics and Astronomy, University of Minnesota,
    Minneapolis, MN 55455}
\altaffiltext{3}{RTN Fellow}
\altaffiltext{4}{Department of Earth Science, Pusan National University,
    Pusan, 609-735 Korea}
\altaffiltext{5}{Department of Astronomy \& Space Science, Chungnam National
    University, Daejeon, 305-764 Korea}
\altaffiltext{6}{Submitted to {\it The Astrophysical Journal}}

\begin{abstract}

We investigate the generation and
distribution of high energy electrons 
in the cosmic structure
environment and their observational consequences by carrying out 
the first cosmological simulation that includes directly
cosmic ray (CR) particles. Starting from cosmological initial conditions,
in addition to the gas and dark matter related quantities,
we follow the evolution of CR electrons (primary and secondary) and CR ions
along with a passive magnetic field.
CR ions and primary electrons are injected in accord to the thermal
leakage model and accelerated in
the test particle limit of diffusive shock acceleration
at shocks associated with large scale
structure formation.
Secondary electrons are continuously generated through p-p inelastic
collisions of the CR ions with the thermal nuclei of the inter-galactic
medium.
The evolution of the CR electrons accounts for spatial transport,
adiabatic expansion/compression, and losses due to Coulomb collisions,
bremsstrahlung, synchrotron and inverse-Compton emission.
The magnetic field is seeded at shocks according to the Biermann battery
model and, thereafter, amplified by shear flow and gas compression.

We compute the emission due to the inverse-Compton scattering of the simulated 
primary and secondary electrons off cosmic microwave background photons
and compare it with the published values of the detected 
radiation excesses in the hard X-ray and extreme ultra-violet wave-bands.
We find that,
from the perspective of cosmic shock energy and acceleration efficiency,
the few instances of detection of 
hard X-ray radiation excess could be explained in the framework
of IC emission from primary electrons in clusters characterized 
by high accretion/merger activity. On the other hand,
with the only exception of 
measured flux from the Coma cluster by \cite{bobeko99}, 
both primary and secondary CR electrons associated with
the cosmic structure formation account at most for a 
small fraction of the radiation excess detected in
the extreme ultra-violet waveband.
 
Next, we calculate the synchrotron emission after normalizing the 
magnetic field strength so that for a Coma-like cluster the volume averaged
$\langle B^2 \rangle ^{1/2}\simeq 3 \mu$G.
Our results indicate that the synchrotron emission from the secondary CR
electrons reproduces several general properties observed in radio halos. 
These include the recently found $P_{1.4GHz}$ vs. $T_x$ relationship,
the morphology and polarization of the emitting region and, to some extent,
even the spectral index. In addition, radio synchrotron 
emission from primary electrons turns out large enough to power
extended regions of radio emission resembling 
radio relics observed at the outskirts of clusters. 
Once again we find striking resemblance between the general properties
of morphology, polarization
and spectral index of our synthetic maps and those 
of reported in the literature.

\end{abstract}



\section{Introduction} \label{intro.se}

The existence of extended regions populated by cosmic ray electrons 
(CRes) in at least some clusters of galaxies has been apparent since the
discovery  more than thirty years ago of diffuse, non-thermal radio 
emissions from the Perseus and Coma clusters \citep{leel61,willson70}.
Their importance as indicators of physical processes in the
clusters and cluster galaxies has grown in recent years as the
number of radio-detected clusters has increased, as reports
have appeared of possible diffuse non-thermal emissions in the 
hard X-ray (HXR) and extreme ultra-violet (EUV) 
bands \citep[\eg][]{lieuetal96a,fufeetal99} and as the evidence has 
mounted for a rich variety of 
highly energetic phenomena in and around clusters that seem capable of
energizing the electrons \citep[\eg][]{sarazin99,ensslin99}. Early on
it was recognized that the radiative lifetimes of CRes
are almost certainly too short and their diffusion too slow for them to 
fill an entire cluster if they are produced by a single point
source \citep{jaffe77}. Still, today, the origin of cluster CRes is not
clear, although many proposals have been made. {\it Our purpose here is
to examine through direct numerical simulations the possibility that
CRes can be explained as a by-product of the very large shocks that
accompany the formation and evolution of the large scale structure
of the Universe.}

\subsection{Radio Observations} \label{raob.se}

The number of successful detections of diffuse radio emission from 
clusters of galaxies has substantially increased in the recent years 
and some cluster statistical properties can now be investigated.
The radio sources are typically featured by low surface brightness and
an emissivity that can be described by a steep power law spectrum 
\citep{hanish82,ebkk98}.
From its spectral properties and, sometimes, polarization signatures the radio
emission is interpreted as synchrotron radiation, implying the presence of 
CRes and magnetic fields
\citep{kkgv89,kkdl90,giova93,deissetal97,gife00}.
Early studies on the origin of the relativistic particles 
conceived of them as originating in discrete radio galaxies \citep{jaffe77,
rephaeli77,rephaeli79,valtaoja84} or normal galaxies. 
More recently, the possibility of their acceleration out of the intra-cluster
medium (ICM) thermal electron pool has been considered \citep{liang99}. 
Models commonly assume a continual energization of the 
relativistic electrons by {\it in situ} first or second order Fermi processes
\citep{jaffe77,robulo93,tribble93b,voahbr96,deissetal97,eiwe99,weei99}.
Alternatively, such electrons could be produced as secondary products of
inelastic p-p collisions of CR ions and the thermal intra-cluster 
nuclei \citep{dennison80,vestrand82,blco99}. 
This idea was put forward primarily to explain the wide-spread distribution 
observed for these electrons, which is in conflict with the fact that 
their cooling time scale is too short for them to propagate throughout
a cluster from a single or even a few discrete sources \citep{dennison80}.

Depending on its observed properties the
diffuse radio emission is further classified as either a
{\it radio halo} when its morphology is regular and typically
centered on and resembling the X-ray emissivity,
or as a {\it radio relic} when it is irregular and
located at the periphery of the cluster. 
The former typically has a low degree of polarization ($< 10\% $).
The degree of polarization in the latter case, when measured, 
turns out typically to be high \citep[$\sim 30\%$; \eg][]{feretti99}. 

Radio halos are usually found in rich clusters with high ICM temperature,
$T\gsim 7$keV, and high X-ray luminosity, $L_x (0.1-2.4\mbox{keV})
\gsim 5\times 10^{44}$erg s$^{-1}$ \citep{fegio96}.
Since it usually extends over a linear size of about 1 \hinv Mpc,
the radio emission appears to be a characteristic of
the whole cluster, rather than being associated with any 
of the individual cluster galaxies \citep{willson70}. 
Signatures of a merging process in these clusters are often emphasized 
and the absence of cooling flows cited 
as demonstrating the connection between radio halos and mergers
\citep[a merger would likely disrupt a cooling flow; \eg][]{feretti99}.
This idea seems attractive, because a merger can provide 
enough energy and the turbulence necessary for 
the amplification of the magnetic field and possibly
acceleration of relativistic particles needed to
power a radio halo \citep{tribble93b}. 
In addition, since mergers are transient events, they would also explain 
why this type of radio source has appeared to be only occasional.
Recently, however, \citet{lhba00} found a tight and steep correlation 
between the radio power emitted at 1.4 GHz and the cluster temperature,
suggesting that the apparent rarity of detections should be attributed 
to observational insensitivity to any but the rarest, most massive clusters.
Those authors also claim that the radio properties of clusters with cooling flows
have not been explored sufficiently as yet. 

On the other hand the clusters hosting radio relic sources
are somewhat less massive and cooler than those related 
to radio halos. Radio relics show no apparent correlation with
merger events \cite[although the two can be coincident; \cf][]{feretti99}
and they are definitely observed in clusters
containing cooling flows \citep{bapili98}.
Radio relics are found both near the cores of clusters and
at their outskirts. The spectra are typically steep, but 
explicit cutoffs are relatively rare even though the 
cooling time of the relativistic electrons is much shorter than 
the age of the cluster. Thus an efficient production mechanism 
for an extended particle population is required in order to explain
the origin of those relativistic electrons, as well. 
In En{\ss}lin's model \citep{ensslin99} a seed population of
relativistic electrons is provided by radio ghosts, \ie
relic relativistic plasma previously injected by radio galaxies.
Such a plasma would be re-energized by the encounter with a shock due either to
a merger event or to supersonic inflow along cosmic filaments \citep{ebkk98}.
Given the high sound speed inside the relic plasma, such flows would
actually become sub-sonic and propagate as a sound wave through the relic,
mostly producing only an adiabatic compression of it \citep{engo01}.

\subsection{Hard X-ray and Extreme Ultra-violet Observations}\label{euvehxr.se}

An energy excess with respect to 
the thermal emission has been detected in 
the HXR band of the spectrum at least in 
four instances: for the Coma cluster
\citep{fufeetal99}, for Abell 2199 \citep{kaastraetal99} and 
for Abell 2256 \citep{fufeetal00}. An upper limit for a HXR non-thermal 
excess was recently measured also for Abell 3667 \citep{fufeetal01}.
Although possibly originating from 
high temperature shocked intra-cluster gas, HXRs are also 
expected as a consequence of 
inverse-Compton (IC) scattering 
off cosmic microwave background photons
by electrons responsible for synchrotron emission.
An approximate ``RMS'' magnitude of the magnetic field can be inferred
from combining HXR and synchrotron surface brightness estimates,
under the assumption of a common
population of relativistic particles generating the two observed radiations.
However, the result for the Coma cluster indicates a value of
$B\sim 0.16 \mu$G \citep{fufeetal99}, which is more than
an order of magnitude lower than magnetic field estimates through Faraday rotation
measures, $B\sim 6 \mu$G \citep{fdgt95} \citep[see also][]{clkrbo00}. An
equipartition estimate of magnetic field and relativistic electrons
results in $B\sim 0.4 \mu$G \citep{giova93}.
There is no good theoretical argument for the existence of that kind of 
equipartition, however.
Given these uncertainties, the origin of the HXRs is still open to debate.
For this reason, non-thermal bremsstrahlung emission has also been proposed
to explain the origin of the HXRs \citep{elb99,sake00,blasi00,dogiel00}.
However, \citet{petro01} argues that, because of the very low efficiency of
non-thermal bremsstrahlung with respect to Coulomb losses, 
in the above model a large amount of energy
would need to be dumped in the ICM, leading to an unacceptably 
large consequent equilibrium temperature.

Additional evidence for the existence of non-thermal populations
of relativistic electrons may be suggested by observations of clusters
in the EUV. 
There have been reported detections of 
EUV radiation from clusters in excess of what is 
expected from the hot, thermal X-ray emitting ICM \citep{lieuetal96a,
lieuetal96b, fabian96, mll98, libomi99, kaastra00, bolimi00}. 
This result appears, however, to be sensitive to the procedure 
adopted for the subtraction of the background
\citep[see, \eg][for the details]{lieuetal96b,mll98,bobeko99}. 
In any case, one
mechanism proposed for the origin of an EUV excess is 
IC scattering off cosmic microwave background photons
of low energy relativistic electrons with a Lorentz factor of
$\gamma \sim 300$ \citep{sali98}.
As an alternative, the possibility that the EUV originates from warm gas 
produced at the interface between a population of cold clouds 
and the hot ICM has also  been explored \citep{fabian96,bolimi00}.
However, observations of the FUV spectra with FUSE has ruled out the existence
of a substantial amount of warm gas, at least, in the Coma and Virgo
clusters \citep{dshl01}.

In summary, there are indications of 
excess, ``non-thermal'' HXR emissions and, possibly
EUV emissions, from some clusters that may signal the presence
of substantial CRe populations there. This information
compliments the radio data, even in its current, uncertain state,
and adds to the motivations for understanding the likely sources and
distributions of CRe populations in clusters.

\subsection{Cosmic Shock Waves}\label{shock.se}

Collisionless plasma shocks are now generally recognized as efficient 
CR accelerators through the so-called ``diffusive shock
acceleration'' (DSA) process \citep[\eg][]{drury83}.
Cosmic structure formation simulations have demonstrated that 
strong accretion shocks form several Mpc's from cluster cores 
and also penetrate deeply into cluster cores
during the course of large-scale structure formation
\citep{kcor94,minetal00}.
They have been suggested as possible sources of high energy CRs
\citep{karyjo96,karabi97}
as well as seeds for the ICM magnetic field \citep{kcor97, rykabi98}.
\cite{minetal00,mrkj01} pointed out that those shocks associated with
large scale structure formation,
in principle have sufficient power to produce dynamically significant CR
proton populations inside clusters, and by 
implication substantial populations of CRes. 
Those computations, based on cosmological simulations, also showed, 
however, that cluster shock structures are very complex.
Thus, it is necessary to explore in some detail the physics of CR
acceleration in this context before one can properly evaluate its role.
In this paper we will investigate by means of a numerical simulation
CRes injection and acceleration by such cosmic shock 
waves. 
In particular, we explore the possibility that the origin of
the aforementioned radio emission and radiation excesses can be explained
by the primary and secondary CRes and focus
our attention on those quantities that are directly related 
to observations. 
These include the flux and the spectral features of the
radiation emitted at various wavelengths, including
the EUV, soft and HXRs produced by IC scattering 
and radio synchrotron emission from the CRes.

There have been previous attempts to study the observational
consequences of CRes with simulations
\citep[\eg][]{robust99,tana00,doen00}. However, for the first time we 
include the CR population {\it explicitly} in a {\it fully cosmological}
simulation by computing particle injection,
acceleration and energy losses in accord with the properties
of the local environment in which the particles are propagating
and starting from cosmological initial conditions.
Here, our focus is CR ``electrons''; CR ``protons'' (as the dominant ionic
CR component) are discussed 
in a companion paper \citep{mrkj01}. 
We note the suggestions that there are additional sources of CRs in clusters,
of course, such as active galaxies \citep{voahbr96,ebkw97,bbp97}.
We do not attempt to include them in our
current work, since our goal is to understand the role of
structure shocks. However, we do call attention in our discussions
to some expected
differences between shock CR sources and isolated sources as appropriate.
Our modeling efforts provide some initial
clues as to how these different sources can be distinguished observationally.

The paper is organized as follows.
In section \ref{num.s} we describe the numerical methods.
In \S \ref{res.s} we present the main results of our simulation
and the conclusions are given in \S \ref{disco.se}.

\section{The Numerical Simulation} \label{num.s}

In order to investigate the various issues related to 
CRes, we have carried out a cosmological 
numerical simulation that follows simultaneously the
evolution of dark ``matter'' particles, gas
quantities, magnetic field, plus CR ions, primary 
CRes injected from thermal plasma and secondary CRes.
The cosmological aspects of the simulation 
setup are presented in the following subsection (\S \ref{num1.se}).
For the matter part, we have used an Eulerian ``TVD'' hydro~+~N-body
cosmology code \citep{rokc93}.
The magnetic field is seeded at shocks
in accord to the Biermann battery mechanism and is thereafter
followed passively \citep{kcor97}. After being generated at
shocks, the magnetic field is amplified by shear flow through 
stretching and by field compression. Numerical resolution in this simulation, however, 
is insufficient to produce field strengths of the same order
of magnitude as observed in today's groups/clusters of galaxies from realistic seed fields.
Therefore, the strength of magnetic field is normalized {\it a posteriori}
to be consistent with the values inferred from 
observations of Faraday rotation measure through 
clusters \citep[\eg][]{clkrbo00}.
Finally, the evolution of the different CR populations in the simulation
is computed by the code COSMOCR \citep{min01}. In the following sections 
we provide a brief description of the physical processes included 
in this code, \ie CR injection at shocks and spatial 
transport (\S \ref{inj.se}), 
production of secondary electrons (\S \ref{prosec.se}) and 
energy losses (\S \ref{radlos.se}).

\subsection{Simulation Set-up} \label{num1.se} 

Since the computation involves a quantity never before simulated
in this context, \ie the CRs, 
we have elected to begin the study from
the relatively simpler case of a standard cold dark matter (SCDM) model,
based a modest resolution simulation with $256^3$ cells and
$128^3$ dark matter (DM) particles.
We leave the currently more favored CDM + $\Lambda $ model 
as the natural follow-up step for future work. 
Although it is now well known that SCDM is not a 
viable model \citep[\eg][]{wnef93,ostrk93},
the key feature required of the simulation is a distribution of
collapsed objects whose properties resemble those observed 
in the real universe,
thus allowing comparisons of their general characteristics.
To accomplish this within SCDM, we set the value for the
root mean square of today's density fluctuation on a scale of
8 \hinv Mpc  to be $\sigma_8 = 0.6$, incompatible with COBE results
and a SCDM universe,
yet a value that induces the emergence of a reasonable population of groups/clusters 
in simulations of large scale structure formation \citep{osce96}.
In addition to $\sigma_8 $ we also define the following parameters:
the spectral index for the initial power spectrum 
of perturbations, $n$ = 1, the normalized Hubble constant, 
$h \equiv $ H$_0 / (100$ km s$^{-1}$ Mpc$^{-1}) = 0.5$, the
total mass density $\Omega_M = 1$ and the baryonic mass 
fraction $\Omega_b = 0.13$. 

The size of our simulated region is chosen as a compromise between a 
simulated cosmological volume large enough to contain groups/clusters
of galaxies and a resolution sufficient to capture the main
structural features of the simulated objects (groups/clusters and their shocks).
We have selected a cubic region of size 50 \hinv Mpc at the current epoch.
With $256^3$cells 
this corresponds to a spatial resolution of about 200\hinv kpc.
In general a coarse grid limits the finest structure
that can form and consequently influences properties of the simulated objects.
Then, because density peaks are smoothed,
estimates of any quantity depending on the square of the density,
such as the thermal bremsstrahlung emission ($\propto n_{thermal}^2$),  
will be reduced. The effect is stronger for 
lower temperature groups/clusters, which have similar structures to the 
larger clusters, but also have smaller physical scales.
Thus, because of resolution effects the 
emissivity from secondary electrons (whose production roughly scales
as $\propto n_{thermal}^2$) will be
systematically underestimated and will have a steeper temperature dependence.
Previous resolution studies carried out to test the performances of the code
employed here indicate that the thermal emission
is underestimated by a factor of a few, although 
the order of magnitude should be correct \citep{ceos99b}.
On the other hand, the resolution errors on the emissivity from primary 
electrons should be less severe compared to the case of secondary 
electrons.

\subsection{Injection at Shocks and Spatial Transport} \label{inj.se}

Ions and electrons injected at shocks from thermal plasma should 
provide the main source of CRs and 
are usually referred to as {\it primary} CRs. 
However, the hadronic interaction of high energy ions
with the thermal nuclei of the background gas 
generates, among other secondary decay products, relativistic 
electrons and positrons. 
The COSMOCR code follows explicitly the injection, acceleration, transport and 
energy losses of three CR species: ions, primary electrons
and secondary electrons. 
The complete CR computational methods are described in \citet{min01},
and the detailed results of the CR ion simulation are presented in 
\citet{mrkj01}.

The injection of CR ions at shocks is computed 
in the ``test particle limit'' and according to a ``thermal 
leakage'' model \citep{kajo95}.
The fraction of supra-thermal protons injected into CRs from plasma flowing
through a shock is controlled by an injection parameter, $c_1$, 
which defines the momentum threshold for injected protons as
$p_{inj} = c_1\,2\sqrt{m_p kT_{sh}}$. 
In this expression
$m_p$ is the proton mass, $k_{\rm B}$ the Boltzmann's constant
and  $T_{sh}$ the post-shock plasma temperature.
We adopted $c_1 = 2.6$, which leads to injection of
$\sim 10^{-4}$ of the thermal protons passing through shocks.
This injection rate is consistent with observation and theoretical
studies of DSA \citep{lee82,quest88,kajo95,malkov98}.
 
Since thermal electrons have gyro-radii much smaller than analogous ionic
gyro-radii, they cannot be injected in the same fashion as
protons. A pre-acceleration process is needed before
electrons can be introduced into the DSA process
\citep[\eg][]{mcclem97}.  
However, the physics underlying the injection at shocks of primary 
electrons is very complex and not well understood. Therefore, 
following the approach of \cite{elbeba00}, we compute the injection 
rate of CRes by assuming a fixed value $R_{e/p}$
for the ratio of primary electrons to protons at relativistic energies. 
There is some observational evidence that this ratio is in the range 1-5 \% 
for the Galactic CRs \citep{muta87,mulletal95}.
Thus, the normalization of the number density in our calculations 
is controlled through the ionic injection rate for secondary electrons
and both the ionic injection rate and $R_{e/p}$ for primary electrons.
For convenience we report our results
with $R_{e/p}=1$. Thus the emissivity for the primary electrons 
needs to be multiplied by an additional factor $R_{e/p}\sim 0.01 - 0.05$
in order to get practical values.

Spatial diffusion of the GeV CRes on the scales of interest
here occurs on time-scales much longer than advection and can therefore
be neglected in regions dominated by smooth flows, \ie shock-less 
regions \citep{jre99,min01}. 
However, the diffusive properties of the CRs are still important because:
(1) the acceleration process depends 
explicitly on diffusive propagation immediately adjacent to the
shock and both (2) the maximum energy reachable by the accelerated particles
and (3) the maximum energy of the particles confined within the ICM depend on 
the diffusion coefficient. 
As for point (1), except for extremely high energy CRs the scales on which that 
diffusion applies
are far too small to be modeled in a cosmological simulation,
and the time scale for the acceleration is effectively
instantaneous by the same measures. 
Therefore, as explained in \cite{jre99},
the injected CR protons and electrons are redistributed
in energy according to a power law as prescribed by DSA
theory \citep[\eg][]{drury83}. Analogously, a pre-existing 
population of CRes passing through a shock
is compressed, and, if
appropriate, its spectrum flattened according to the Mach number of the shock.
Additional details of these procedures can be found in \cite{min01}.
As for points (2) and (3), the diffusion coefficient and the 
associated mean free path are usually computed in the pitch 
angle resonant scattering approximation \citep[\eg][]{drury83}
as 
\begin{equation}
\kappa_\parallel =\frac{1}{3}
r_L v \frac{B^2/8\pi}{k W_k},
\end{equation}
\noindent
where $W_k$ is the magnetic energy density per wavenumber 
evaluated at $k\sim 1/r_L$, \ie the inverse of the Larmor radius. 
Thus the diffusion coefficient 
is inversely proportional to the level of resonant wave turbulence,
and depends, therefore, on the spectral properties 
of the turbulent magnetic field.
The diffusion coefficient determines the rate of energy gain which, when 
compared with the dominant energy loss rate gives an estimate of 
the maximum energy of the accelerated particles. 
It is shown that in the 
case of an average ICM magnetic field of value $\sim 0.15 \mu$G and with 
a Kolmogorov power spectrum, the upper limit for the 
energy of the accelerated electrons could be low enough to 
produce a cut-off in the radio 
synchrotron emission around a few GHz \citep{blasi01}. 
Such a cut-off, however, disappears if one makes the assumption of 
Bohm diffusion \citep{blasi01}. 
In any case, with the normalization adopted in \S \ref{radlos.se} 
the magnetic field in the groups/clusters of interest here 
(with temperature $T_x\gtrsim 1$keV) is always larger
than the above value of 0.15 $\mu$G.
Therefore, such a cut-off should not affect
the spectra of our simulated clusters. 
Finally, we point out that diffusive escape of 
the CR-ion component, which is responsible for the 
production of the secondary electrons, only affects particles 
with energy above $\sim 2\times 10^{15}$ eV \citep{voahbr96,bbp97}. 
This component was not included in our
simulation \citep{mrkj01}
and should not be relevant for the secondary electrons in the 
energy range investigated here \citep[\eg][]{masc94}.

\subsection{Production of Secondary Electrons} \label{prosec.se}

The main channels for the production of secondary electrons
are given by \citep{gaisser90}:
\begin{eqnarray} 
p + p & \rightarrow & \pi^\pm + X  \label{pppi.eq}\\
p + p & \rightarrow & K^\pm + X, \label{ppk.eq}
\end{eqnarray}
\noindent
where $X$ indicates all the other by-products of the reactions.
Subsequently charged pions and kaons decay mainly into 
muons (although a fraction of kaons also produce pions), 
from which the secondary electrons originate 
\citep[see][for more details on the processes included in COSMOCR]{min01}.
In addition to $p+p$ inelastic collisions, 
the above cascades are also triggered by the interaction of
$p+$He, $\alpha+$H and $\alpha+$He which, for example, increase the
the overall yield of secondary $e^-$
by a factor 1.4 
for a metal composition similar to the inter-stellar 
medium \citep{dermer86}. 
Here we have assumed a helium number fraction
of 7.3\% for the background gas and a ratio
$(H/He) \simeq 15$ at fixed energy-per-nucleon 
for the CRs \citep{medrel97}.

In general we write the production spectrum of
secondary electrons as \citep{most98}:
\begin{equation}  \label{emissnt.eq}
i_s(\varepsilon_s) = n_H \: \sum_{i = \pi, K} 
\int_{\varepsilon_p^{min}}^\infty
d\varepsilon_p \; J_p(\varepsilon_p)\, 
\langle \zeta\sigma_i(\varepsilon_p) \rangle \:
\int_{\varepsilon_i^{min}(\varepsilon_s)}^{\varepsilon_i^{max}(\varepsilon_p)}
d\varepsilon_i \: F_s(\varepsilon_s, \varepsilon_i) 
F_i(\varepsilon_i, \varepsilon_p),
\end{equation}
\noindent
where $J_p(\varepsilon_p) $ is the proton flux; 
$\langle \zeta\sigma_i(\varepsilon_p) \rangle$ is the inclusive cross
section\footnote{Inclusive means that which describes the process
$ p+p \rightarrow i + X$, where $i$ is in general a secondary particle.}
of the processes \ref{pppi.eq} and \ref{ppk.eq};
$\varepsilon_p^{min}$ is the minimum proton energy required to 
produce a meson of energy $\varepsilon_i^{min}$ and $\varepsilon_i^{max}$
the maximal energy of the produced meson;
$\varepsilon_i^{min}$ in turn is the minimum required energy of a meson 
for production of secondaries of energy $\varepsilon_s$; finally 
$F_i(\varepsilon_{\pi,K}, \varepsilon_p)$ 
are the spectra of $\pi$ and $K$ produced from the collision of a proton of
energy $\varepsilon_p$
and $F_s(\varepsilon_s, \varepsilon_{\pi,K})$ the distribution of secondaries 
from the subsequent decay of the above collision products.
The cross sections were computed according to the GALPROP routines
by \citet{most98} to which we refer for  
a full summary of the technique employed here for the calculation
of the secondary electrons.

\subsection{Radiative Losses} \label{radlos.se}

COSMOCR \citep{min01} accounts for various energy loss mechanisms suffered
by the CRes in different energy ranges of interest here.
To do this, a Fokker-Planck equation 
that has been integrated over finite momentum bins 
is solved. 
This takes advantage of the
near-power-law form of the CR momentum distribution.
In effect, the momentum space is divided into $N_p=
8$ logarithmically equidistant intervals, bounded by 
$ \phat_0, ...\phat_{N_p}= \phat_{max}$, 
which we refer to here as {\it momentum bins}.
The electron distribution function $f(\hat{p})$,
as a function of the normalized momentum $\hat{p}\equiv p/m_ec$,
in each spatial cell and for each momentum bin
is approximated by the following piece-wise power law:
\begin{equation} \label{distf.eq}
f({\bf x}_i,\phat) = f_j({\bf x}_i) \, \phat^{-q_j({\bf x}_i)},
~~~~~~~~ 1<\phat_{j-1} \le \phat \le \phat_j,
\end{equation}
where $f_j({\bf x}_i)$ and $q_j({\bf x}_i)$ are the number normalization 
and logarithmic slope 
for a given cell and $\hat{p}$ bin.
With the above definition, the number density of particles
is given by $dN = 4\,\pi\,\hat{p}^2\, f(\hat{p}) d \hat{p}$. 

Within each momentum bin, $j$, and at each spatial grid point, ${\bf x}_i$,
we follow the CRes total number density and kinetic energy density defined as 
\begin{eqnarray}
n({\bf x}_i,\phat_{j}) & = &  4\pi\;
\int_{\phat_{j}}^{\phat_{j+1}} f({\bf x}_i,\phat) \phat^2 d\phat \\
\varepsilon({\bf x}_i,\phat_{j}) & = &  4\pi\;
\int_{\phat_{j}}^{\phat_{j+1}} f({\bf x}_i,\phat) T(\phat) \phat^2 d\phat,
\end{eqnarray}
where $T(\phat)=(\gamma-1) m_ec^2$ is the relativistic 
kinetic energy.
Further, for each momentum bin
$q_j({\bf x}_i)$ is determined self-consistently 
from the values of 
$n({\bf x}_i,\phat_j)$ and $\varepsilon({\bf x}_i,\phat_j)$
defined above \cite[see][for details]{min01}.
With this formalism, the evolution of $n({\bf x}_i,\phat_{j})$ 
in momentum space is described by the equation
\begin{equation} 
\frac{\partial n({\bf x}_i,\phat_{j})}{\partial t} 
= - {\bf \nabla\cdot [ u}\,n({\bf x}_i,\phat_{j})]  
+ \left[ 
b(\phat) \;4\pi \;\phat^2\; f(\phat)\right] _{\phat_{j-1}}^{\phat_j} 
+ Q({\bf x}_i,\phat_{j}) \label{dce3.eq}
\end{equation}
where the first term on the right hand side 
describes advective transport and $Q({\bf x}_i,\phat_{j}) $
represents the source term, $i({\bf x}_i,\phat)$, 
integrated in the $j_{th}$ bin.
Finally, $b(\phat)\equiv dp/dt$ describes
mechanical and radiative loss terms \cite[see][for further details
on this]{min01}. 
The most effective among these are 
Coulomb losses in the low energy end and synchrotron and IC
emission at the highest energy. 
Bremsstrahlung losses are less important, although included for completeness.
Losses due to Coulomb collisions are \citep{stmo98}
\begin{eqnarray}  \label{coul.eq} \nonumber
\left(\frac{d\phat}{dt}\right)_{Coul} & = &
1.1\times 10^{-12} \left\{1+[\ln(1+\phat^2)^{1/2}-\ln n]\frac{1}{73.56}
\right\}\;n
\;\mbox{s}^{-1},
\end{eqnarray}
where $n$ is the number density of the background
gas in cm$^{-3}$.
Bremsstrahlung losses are defined as 
\citep{ginzburg79,stmo98}
\begin{eqnarray} \label{bremm.eq} \nonumber
\left(\frac{d\phat}{dt}\right)_{brem} & = &
1.4 \times 10^{-16}\left\{\ln[2\,(1+\phat ^2)^{1/2}
]\,-\frac{1}{3}\right\}\, \phat\,n
\;\mbox{s}^{-1}.
\end{eqnarray}
For synchrotron emission and IC scattering,
the combined contribution is given by
\begin{eqnarray} \label{syncic.eq} \nonumber
\left(\frac{d\phat}{dt}\right)_{sync+IC} & = &
 1.36\times 10^{-20} \left(1+\frac{u_B}{u_{cmb}}\right)
\phat^2 \;\mbox{s}^{-1},
\end{eqnarray}
where $u_B$ is the energy density in magnetic field, and
$u_{cmb} = 4.2\times 10^{-13} \,(1+z)^4\;{\rm erg ~cm^{-3}}$ 
is the energy density in cosmic microwave background.

Since our simulation treats the magnetic field passively, 
it is necessary to normalize the magnetic field strength.
For this purpose, we assume that the volume
averaged magnetic field in a Coma-like cluster of $\sim 8$ keV is of order 
3 $\mu$G, which corresponds to values inferred from
Faraday rotation measures \citep{kkgv89,kkdl90}.
Thus for most of the groups/clusters in our 
simulation, characterized by a temperature $T_x \lesssim 3$keV, 
the energy density in the magnetic field is
smaller than that in the cosmic microwave background and 
the radiative loses due to IC process dominate over synchrotron
losses. 
The IC cooling time of CRes is of order
\begin{equation} \label{icct.eq}
\tau_{ic} = 2.3\times 10^8 \left(\frac{\phat}{10^4}\right)^{-1} ~~{\rm yr},
\end{equation}
where $\phat = \gamma \sim 300~(10^4)$ is required for electrons generating IC 
emission in the EUV (HXR) band.

\subsection{Calculation of Emissivity} \label{calce.se}

In this section we described the procedure adopted for the calculation
of IC emission in the HXR and EUV bands as well as radio synchrotron at
various wavelengths of interest. 

\subsubsection{Inverse-Compton Emission}

IC emission is computed 
in units of erg cm$^{-3}$ s$^{-1}$ Hz$^{-1}$ sr$^{-1}$ 
according to
\citep[\eg][but note the different units]{rpa79} 
\begin{equation}
j_{ic}(\nu)  =  \frac{3\, c \,\sigma_T}{4}
h \nu  \,\int v_{cmb} (\nu_0) \, d\ln\nu_0 
\sum_{j=1}^{N_p} f_j
\,\int_{\phat_{j-1}} ^{\phat_j}
\; g\left(\frac{\nu}{4\hat{p}^2 \nu_0}\right) \,
\hat{p}^{-q_j} \; d\hat{p} 
\label{icepl.eq}
\end{equation}
where $g(x)=2x\ln x-2x^2+x+1$,
$v_{cmb} (\nu_0)= u_{cmb} (\nu_0)/h\nu_0$ 
indicates the number density of photons in the 
radiation field at frequency $\nu_0$ and the sum extends over all the bins
of the simulated electron populations defined in \ref{distf.eq}.
The above equation \ref{icepl.eq} is the actual formula 
that we used to calculate the inverse Compton
emission at various wavelengths from the simulated electrons.
Since most of the emission is contributed from the CRes in the 
$k_{th}$ bin such that $\phat_{k-1} \le (\nu/4 \nu_0)^{1/2}\le \phat_k$, 
for the sake of illustration we also write
\citep{rpa79}
\begin{equation}
j_{ic}(\nu)  \simeq   f_k
\frac{8 \, \pi^2\,h \,r_0^2}{c^2} 
\left(\frac{kT_{cmb}}{h}\right)^{(q_k+3)/2} \, L(q_k-2) \;
\nu^{-(q_k-3)/2} .
\label{icepl2.eq}
\end{equation}
where
\begin{eqnarray}
L(q) = 2^{q+3}  \,
\frac{q^2+4q +11}{(q+3)^2 (q+5)(q+1)} \, \Gamma\left(\frac{q+5}{2}\right)
\sum _{n=1}^\infty n ^{-(q+5)/2}.
\end{eqnarray}
Thus from eq. \ref{icepl2.eq} we infer that the
frequency dependence of the IC
emissivity is a power-law, $j(\nu) \propto \nu^{-\alpha}$,
with {\it spectral index}
\begin{equation}  \label{spidx.eq}
\alpha \simeq \frac{q_k-3}{2}.
\end{equation}

\subsubsection{Synchrotron Radiation}

Analogously, for a piece-wise power-law distribution of electrons given 
in eq. \ref{distf.eq}, the synchrotron emissivity
(in units of erg cm$^{-3}$ s$^{-1}$ Hz$^{-1}$ sr$^{-1}$)
is computed according to the expression
\begin{equation} \label{jsync.eq}
j_{syn} = \frac{\sqrt{3} e^3 B}{2 m_ec^2} 
\; 
\sum_{j=1}^{N_p} f_j \;p_j^{q_j}\;
\left(\frac{2\nu}{3\nu_B}\right)^{-(q_j-3)/2}
\int_{x_{j-1}} ^{x_j}
F(x)\,x^{-(q_j-5)/2} \;dx
\end{equation}
where $\nu_B = eB/(2\pi m_e c)$, $x_j=1.5 \,\nu/\nu_B\, \phat^{-2}_j$ 
and we approximate $F(x)=0.87\, x^{1/3}\, \exp{(-11\,x^{7/8}/8)}$
\citep{elb99}.
As for the inverse Compton emission we can 
also derive an approximate expression \citep{gisy65}
\begin{equation} \label{jsync2.eq}
j_{syn} \simeq f_k \, \frac{\sqrt{3} \, e^2\, \nu_B}{2 \, c (q_k-1)}
\, \Gamma\left(\frac{q_k}{4}+\frac{13}{12} \right)\, 
\Gamma\left(\frac{q_k}{4}-\frac{7}{12} \right)
\, \left(\frac{2\nu}{3\nu_B}\right) ^{-(q_k-3)/2},
\end{equation}
with $k$ referring to the momentum bin such 
that  $\phat_{k-1}\le \sqrt{4.5\, \nu/\nu_B} \le \phat_k$.
Again, from \ref{jsync2.eq} we infer a spectral index $\alpha \simeq (q_k-3)/2$, 
as for the IC emission case. The local magnetic field strength, entering $\nu_B$,
is  given by the simulated value after a global re-normalization is performed according
to the criterion given in \S \ref{radlos.se}.
Further, by using the morphology of the simulated field, the degree of polarization
of the synchrotron radiation is also computed.

\subsection{Data Analysis} \label{globp.se}

From the simulation results collapsed objects have been
identified by the DM-based ``spherical over-density'' method
described in Lacey \& Cole (1994).
The details of our identification procedure can be found
in \citet{minetal00}.
Global quantities, $Q$, (\eg luminosity, photon flux, spectral index),
are extracted by integrating or averaging over the group/cluster volume as
\begin{equation} 
 Q = \sum_i q_i \, w_i
\end{equation}
where $q_i$ is the cell value of the quantity of interest and $w_i$ is
the weight function, \ie the portion of 
computational cell within $V=4 \pi\,R^3_{cl}/3$. Values of
$w_i$ have been computed up to a diameter of 13 computational
cell corresponding, given our current resolution, to 
radius $R_{cl}\simeq$ 1.3 \hinv Mpc.
Some of these global quantities are studied as a function 
of the group/cluster core temperature, $T_x$. 
The latter is defined as the average
temperature within a central region of diameter 0.5 \hinv Mpc,
the typical size of the observed X-ray clusters (and over
which the temperature is observationally determined).
A temperature dependence is evaluated 
in general by fitting a computed global 
quantity, $Q$, to the following power-law,
\begin{equation} \label{fit.eq}
 Q = K \; \left(\frac{T_x}{6.72\mbox{keV}}\right)^{\phi} .
\end{equation}
where $K$ and $\phi$ are the fitting parameters.
The best fit parameters are then retrieved by 
a least $\chi^2$ analysis.
Global quantities such as the luminosity at different 
wavelengths are expected to correlate with $T_x$ since the 
emitted radiation should depend on the available group/cluster
gravitational energy. In any case, we use
the values of the scaling parameters
to make predictions about the quantities under investigation and
to make comparisons between the 
predictions from our simulation and observations.

Alternatively, two-dimensional maps have been constructed from 
the data-set, either simply as slices through the volume of physical 
a physical quantity (\eg density); or by 
way of a line of sight integration of a volume emissivity 
relative to a selected simulated group/cluster.
The latter procedure produces synthetic surface brightness
maps in the optically thin 
approximation through a projection code \citep{iant01}. 
For comparison purposes the groups/clusters are
set to a luminosity distance corresponding to the red-shift of the 
Coma cluster ($z=0.023$), \ie about 
70 \hinv Mpc. 
Since our resolution amounts to $\sim$ 200 \hinv kpc per 
computational cell, at this distance the minimal size of a pixel 
of the synthetic image corresponds to 9.8 arcmin squared. 
These images 
allow a more in-depth inspection of the spatial distribution of
the quantities of interest, but for practical reasons are limited
so far to only a few samples.

\section{Results} \label{res.s}

\subsection{Inverse-Compton} \label{resic.se}

IC emission from the simulated CRes is computed in the following wave-bands:
in HXR, between 20 and 80 keV, where measurements have
been carried out for Coma cluster 
\citep{fufeetal99}, Abell 2256 \citep{fufeetal00} and Abell 3667 \citep{fufeetal01}, and 
between 0.13 to 100 keV, relevant for Abell 2199 \citep{kaastraetal99}; 
and in the EUV between 
65-248 eV, the wave-band of the EUV Explorer satellite.

Surface brightness maps 
for one of the collapsed objects generated in the simulation
are presented in Fig. \ref{HXR2.f} 
for the HXR emission between 20 and 80 keV in units $2\times 10^{-4}$
erg cm$^{-2}$ s$^{-1}$ per pixel and 
in Fig. \ref{euv2d1.f} for the EUV emission in units $4\times 10^{-7}$
erg cm$^{-2}$ s$^{-1}$ per pixel, respectively.
In each figure, left and right panels correspond to 
secondary and primary electrons, respectively.
From Figs. \ref{HXR2.f} and \ref{euv2d1.f} it appears that 
the radiation from primary electrons is more extended, less
centrally peaked and morphologically more irregular than
that from secondary electrons. 
The irregularity of
the maps corresponding to the primary electrons 
is caused by the fact that latter are preferentially 
found around existing shocks, where they are freshly injected
and have not yet been depleted by radiative losses.
Such irregularity is stronger 
for HXR than for EUV photon energies. 
In fact, primary HXR emitting electrons, with Lorentz factors
$\gamma \sim 10^4$ (\cf eq. \ref{icct.eq}), 
have very short cooling times and do not propagate
far from the shock where they have been generated. 
For the primary EUV emitting electrons, the cooling time is 
about 100 times longer, and in fact these electrons 
are distributed throughout the collapsed object, 
although they are still more
numerous around shocks.
The distribution of secondary 
electrons, on the other hand, and the ensuing morphology of 
IC emission maps, are much less affected by, but not totally insensitive to,
the presence of shocks. 
Their emissivity roughly varies as $n_{gas}^2$, indicating
that although spread more widely, the distribution of the parent CR protons
follows more or less that of gas \citep{mrkj01}.
Such general properties
of the emission from primary and secondary electrons will characterize
the radio maps as well. 

\begin{figure}
\plotone{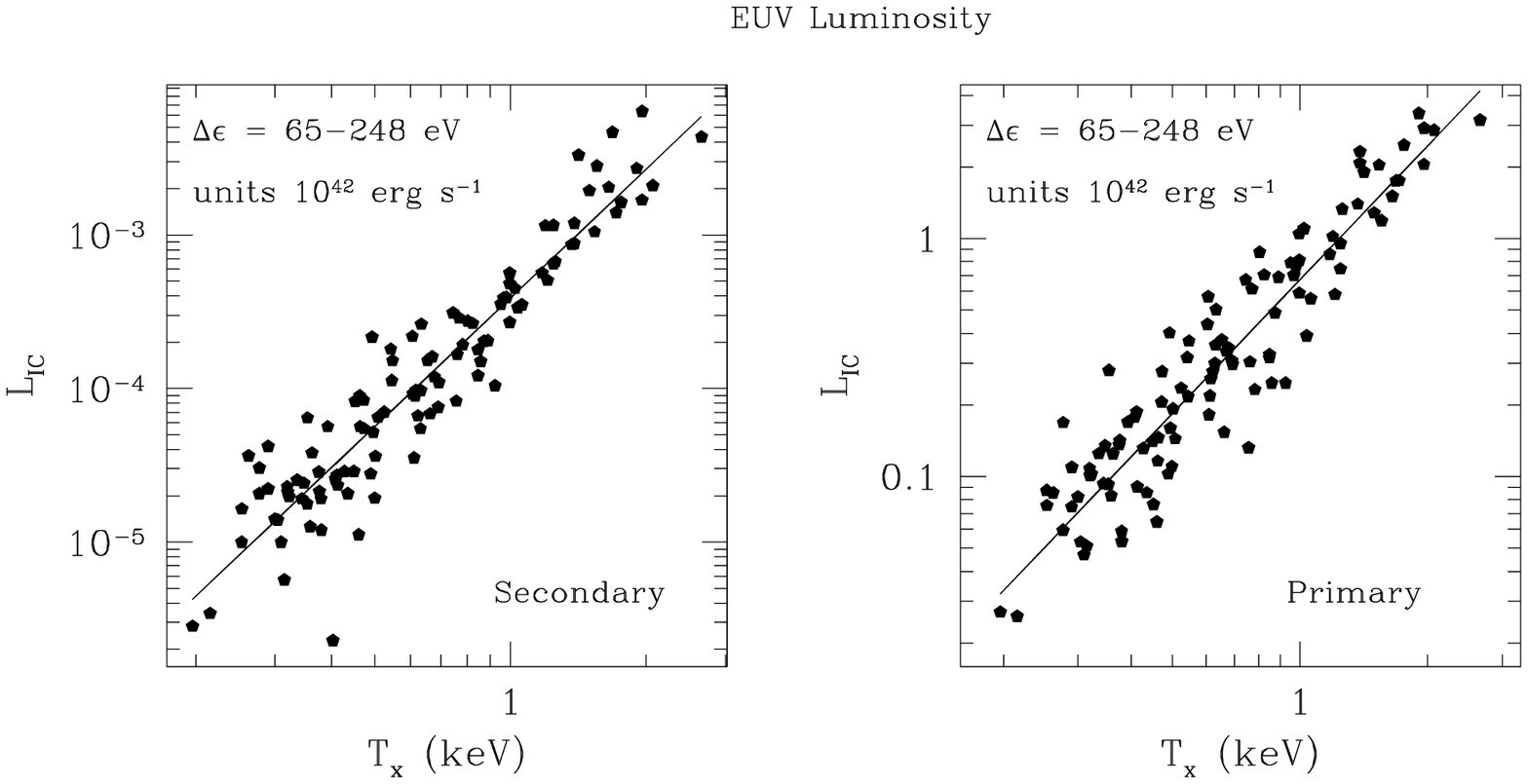}
\figcaption{Group/cluster EUV luminosity from IC
emission due to secondary electrons (left) and
primary electrons (right) as a function of the group/cluster core temperature. 
The emissivity is integrated over a spherical volume 
within a radius of 1.3 \hinv Mpc.
Note that the physical values of the luminosities for
primary electrons should be scaled with the electron to proton 
injection ratio, $R_{e/p}$.  
\label{euvic.f}} 
\end{figure}

The total IC emission in the wave-bands of interest
has been computed for each simulated 
group/cluster and plotted as a function of its core temperature, $T_x$.
In Fig. \ref{HXRic.f}
we report the fluxes of HXR radiation 
in the ranges 0.13-100 keV (left panels)
and 20-80 keV (right panels) 
produced by secondary (top) and primary (bottom) electrons
respectively, for a luminosity distance of the simulated groups/clusters
of 70 \hinv Mpc. 
In Figs. \ref{euvic.f} we present the
luminosity in the EUV band (65-248 eV)
for secondary (left) and primary (right) electrons respectively.
The values reported for the primary electrons assume
$R_{e/p} =1 $ (see \S \ref{inj.se} for further detail).
The data plots have been fitted to a power-law correlation as 
described in \S \ref{globp.se}. 
The best fit parameters are given in Table \ref{fit.tab}.
Note that with primary electrons the slope $\phi$ for 20-80 keV HXR is
somewhat smaller than those for 0.13-100 keV HXR and EUV. This is at least
partly because of larger scatter seen in the bottom-right panel of
Fig. \ref{HXRic.f}. We believe that this scatter has a physical 
origin in the different {\it shock acceleration history}
undergone by each collapsed object identified in the simulation box.
However, we do expect larger 
simulations, which allow both more numerous and larger 
clusters as well as higher resolution, to result in
more consistent values of $\phi$.

The temperature scaling of the emitted radiation
from the two CRe populations is quite different.
This fact corresponds to differences in the 
scaling properties of secondary and primary electrons with the 
group/cluster temperature. 
The scaling of the IC flux from secondaries 
is expected to go as
\begin{equation} \label{ficsl.eq}
F_{ic}(\nu) \propto N_{es} (\gamma _\nu) 
\propto N_{cr}(\gamma _\nu) \,n_{gas},
\end{equation}
where $N_{es}(\gamma_\nu)$ and $N_{cr}(\gamma_\nu) $ 
are the total number 
of secondary electrons with energy $\gamma_\nu$
responsible for the emission at the specified frequency, $\nu$,
and that of the parent CR ions respectively.
By accounting for the gas accretion 
rate and for the injection mechanism adopted here as
roughly converting a fixed fraction of the shocked gas into CRs 
(see \S \ref{inj.se}), it can be shown that
$N_{cr}(\gamma_\nu) \propto T_x^2$.
In addition, the average baryon density inside the group/cluster volume within
the fixed radius of 1.3 \hinv Mpc has an almost linear dependence on $T_x$,
in accord with observations \citep{edst91b,mohr99}.
Although resolution effects discussed in \S \ref{num1.se}
may cause some steepening of the $F_{ic}-T_x$ relation, our
numerical results are consistent with the above arguments.
On the other hand, for the IC emission
from primary electrons we have 
\begin{equation}
F_{ic}(\nu) \propto N_{ep}(\gamma_\nu) \sim N_{cr}(\gamma_\nu),
\end{equation}
since the total number of primary electrons and ions should be 
proportional at the time of injection.
Therefore, in this case the temperature dependence of the flux from
IC scattering should be weaker by roughly one power of $T_x$ with 
respect to the secondary CRes model, as indeed we have found.

\begin{figure}
\plotone{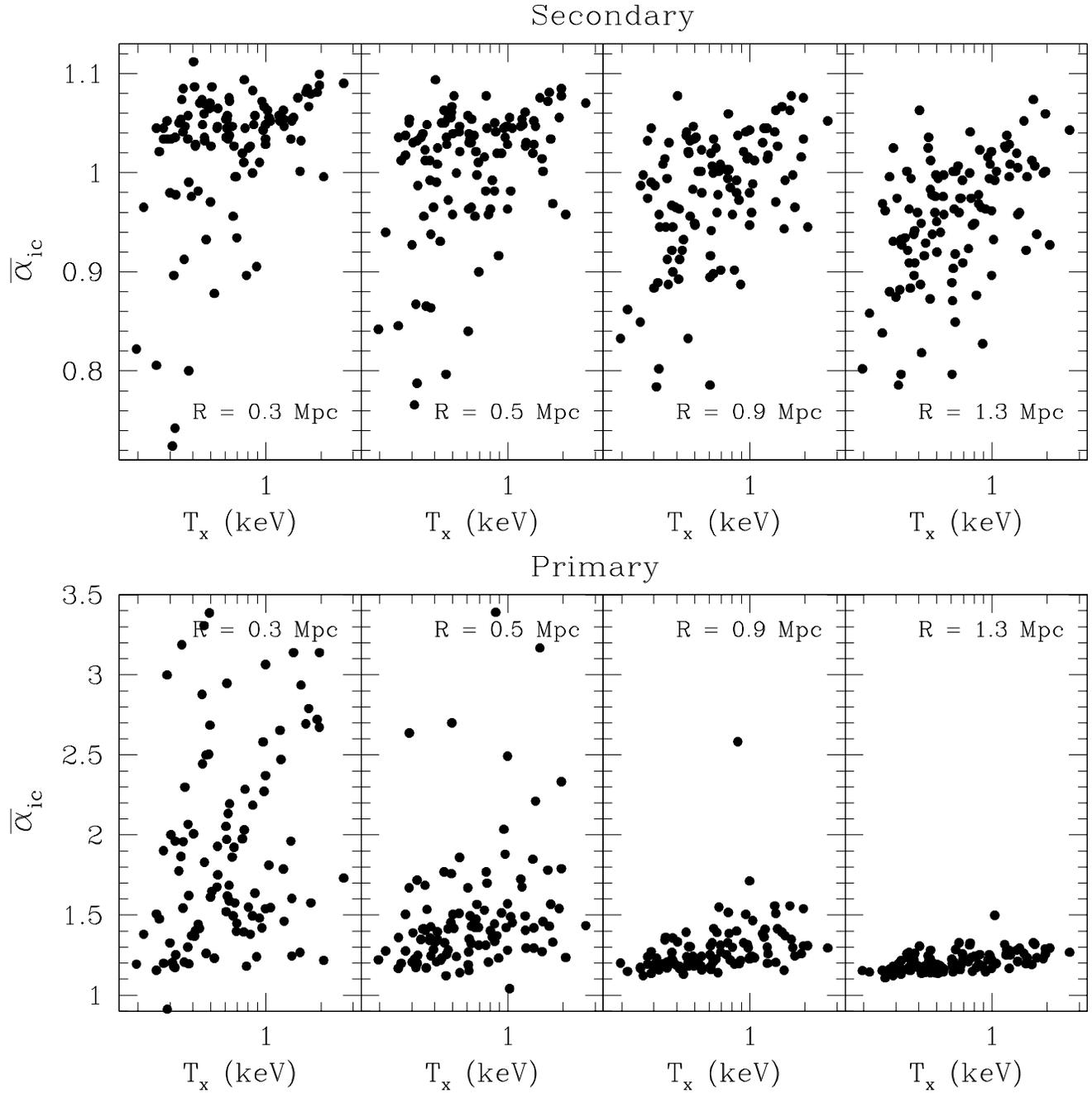}
\figcaption{Top: Volume-averaged spectral index for the HXR radiation
at 50 keV from secondary electrons as a function of the group/cluster core
temperature for four different averaging volume radii, \ie from left
to right $R= 0.3, 0.5, 0.9, 1.3$ \hinv Mpc.
Bottom: same as top but now for primary electrons.
\label{qHXR50.f}} 
\end{figure}

We have also studied the spectral properties of the radiation
emitted by the simulated electrons.
Since the published data report observed values of the spectral index 
for the HXR emission only, we only present our results
for this wave-band, although the same general discussion 
described below applies to EUV emitting electrons too.
Thus, for each simulated group/cluster we have computed
the IC HXR spectral index, $\alpha_{ic}$, 
averaged over a cluster-centered volume defined by a radius $R$ 
to give $\bar{\alpha}_{ic}(<R)$. 
Our results are reported in Fig. \ref{qHXR50.f} for both 
primary (bottom panels) and secondary electrons (top panels),
as a function of the clusters/groups temperature,
$T_x$, for four different values of $R$.

As for the primary electrons the situation 
is complicated because our spatial resolution is inadequate to 
follow in detail the spectral evolution of
this population of CRes immediately downstream of shocks.
In fact, after being shock-accelerated,
CRes will rapidly radiate away their energy as they
are advected away from shock acceleration regions. Thus,
in a realistic picture, at various distances from the shock 
we should find populations of CRes with increasingly ``older age''. 
However, for initial energies of interest the time-scale for cooling 
is typically shorter compared to the time for advection across a numerical cell. 
Therefore, the proper CRes spatial distribution cannot be reproduced 
because of resolution limits. Nevertheless we know that for the case
of IC and synchrotron cooling, the summation 
of all of these ``aging'' populations in the post-shock region produces
a {\it volume-averaged} population with a power-law index steeper
by one from that produced at the shock. 
Thus, the spectral index of the volume-averaged emission from 
the post-shock region can be recovered from the properties of the shock.
In addition, in order to compute correctly the total emission
from the post-shock region,
we adopt a sub-cycling procedure that allows us to estimate 
the steady-state number density of high energy electrons there.
 
With this clarification, in the four bottom 
panels of Fig. \ref{qHXR50.f} we present the volume-averaged 
spectral index associated with the post-shock IC emission region.
According to our discussion above,
the spectral index characterizing the IC emission from 
the high energy electrons that are emerging from the
acceleration region, \ie those immediately behind the shock
is flatter by about 0.5 (\cf eq. \ref{spidx.eq}).
For shocks with Mach number $\ge 3$,
representing most of the shocks in our simulation,
the spectral index is $\alpha_{ic} \le 1.3$.
Indeed, in the last panel (bottom-right) of 
Fig. \ref{qHXR50.f}, corresponding to an averaging volume
of radius 1.3 \hinv Mpc around the group/cluster center
we have $\bar{\alpha}_{ic}\simeq 1.-1.3$. This
volume is large enough to include most of the shocks where 
fresh particles are injected. On the other hand,
for smaller radii, most of the considered volume has been
depleted of energetic particles through
radiative losses. Therefore, we find 
higher values and larger fluctuations for 
$\bar{\alpha}_{ic}(<R)$.

Turning now our attention to the top panels of Fig. \ref{qHXR50.f}, 
we notice that for the secondary electrons 
$\bar{\alpha}_{ic}(<R)$ mostly ranges between 0.8 and 1.1.
The distribution of secondary electrons,$f_{es}$, is determined by
the loss equation, which reads
\begin{equation}
\frac{\partial}{\partial t} f_{es}(\phat) = \frac{1}{\phat^2} 
\frac{\partial}{\partial \phat}  [\phat^2 f_{es}(\phat)\, b(\phat)]
+ i(\phat),
\end{equation}
where $\phat$ and $b(\phat)$ have been previously defined and
$i(\phat)$ is the term describing the source of particles.
For a steady state solution, if $i(\phat)\propto \phat^{-r}$
and $b(\phat)\propto \phat^{t}$, then the particle distribution is
\begin{equation}
f_{es}(\phat) \propto \phat^{-r+1-t}.
\end{equation}
We point out that most of the CR ions
produced in our simulation
have a spectrum characterized by a $q_i=4.0-4.2$ 
\citep{mrkj01}, which translates into a slope
for the source function of secondary electrons close
to the same value, \ie $r\simeq 4.2$.
For IC losses, relevant for the HXR emitting electrons, $t=2$;
thus from eq. \ref{spidx.eq}
\begin{equation}
\alpha_{ic} =  \frac{(r-1+t) -3}{2} \simeq 1.1,
\end{equation}
in rough accord with the values in the top plots of Fig. \ref{qHXR50.f}.
We notice, however, in those plots the presence of several points
indicating values of $\bar{\alpha}_{ic}(<R) < 1$. In addition,
the number of such points becomes more numerous as we 
move toward the right panel, \ie larger averaging volumes 
(Fig. \ref{qHXR50.f}).
These values of the spectral index would imply the presence of
secondary electrons with distributions, $f_{es}(p)\propto p^{-q_{es}}$,
flatter than $q_{es}=r-1+t=4-1+2=5$, \ie than our simulated CR ions 
(with $q_i\simeq 4$) can generate. Such distributions are in fact produced 
by the re-acceleration that some secondary electrons undergo 
as they encounter a shock wave in the ICM. 
This event is more likely to happen at the outskirts 
of a cluster where shocks are more frequent,
in accord with the trend of the top panels in Fig. \ref{qHXR50.f}.
However, since we expect such re-accelerated secondary CRes to 
behave exactly as the primary electrons discussed above, 
the flattening of the spectral index should be only ``visible'' 
very close to the shocks and should disappear when the radiation 
from the large post-shock emission region is considered.
Thus, due to the same coarse grid effects described above
in the context of primary electrons, 
the spectral indexes reported in the top panels of Fig. \ref{qHXR50.f}
tend to be smaller than they should.
Unlike the case of the primary electrons, however, 
correcting for this effect is highly impractical because
it would require following (numerically) the re-accelerated 
particles as an independent population. In any case, 
we expect the values of $\bar{\alpha}_{ic}(<R)$ 
reported in the panel corresponding to an averaging volume of
$R=$ 0.3 Mpc to be only little affected by the re-accelerated 
populations of secondary CRes and, therefore, fairly 
representative of the simulation results.

Finally we point out that for EUV emitting secondary electrons,
with Lorentz factor $\gamma\sim 300$,
the loss mechanism are not only IC emission but also 
Coulomb collisions,
implying a smaller value for both $t$ and $\alpha$.

\subsubsection{Hard X-ray Emission} \label{hxr.se}

Having computed the fitting parameters summarized in Table \ref{fit.tab}
we can now estimate the HXR fluxes for Coma, Abell 2199 and Abell 2256
clusters as predicted by our simulation model.
For this exercise, an extrapolation has been applied using the parameters in
Table \ref{fit.tab}, since the collapsed objects identified in simulation data
have temperature $\la$ 3 keV while the observed clusters used for
comparison have higher temperature.
In order to define the scope of such extrapolations,
first of all we notice that the the plots reported in 
Table \ref{fit.tab} show quite a bit of (real) scatter,
which we attribute to the different evolution of the collapsed objects.
Thus our plots will only allow an estimate of average values and
within a factor of the order of the scatter.
As pointed out above, the scatter in the bottom-right panel of
Fig. \ref{HXRic.f} is particularly large. 
With this clarification, in Table \ref{hxrmp.tab}
we report the name of each observed cluster in the first column,
and the measured red-shift, temperature, HXR flux and 
spectral index in columns 2-5 respectively (references in column 10). 
The expected average HXR flux and spectral index according 
to our secondary model are reported in columns 6 and 7
respectively. Finally, column 8 and 9 refer to the primary CRes model:
the latter lists the spectral index associated with each
population of CR particles and the former 
refers to the required $R_{e/p}$ value in order
for the HXR flux from primary CRes to 
match the reported observations.

The values reported in Table \ref{hxrmp.tab} indicate
that the total number of secondary electrons 
is too small to account for the full
measured HXR flux from any of Coma, Abell 2256 or Abell 2199.
In fact, the calculated emission from secondary electrons in our simulation
corresponds at most to a few percent of the reported measurements.
Higher values of the total injection efficiency 
$\eta_{inj}$ (see \S \ref{inj.se}) could 
allow a larger population of CR ions and, therefore, of 
secondary electrons. However, the resulting 
non-thermal pressure and $\gamma$-ray 
flux associated with the CR ions would then become inconsistent
with the observational limits \citep{mrkj01,blco99,blasi99}.

On the other hand, in order
to account for reported detections in the
HXR wave-band with 
the emission from the primary electrons
in our simulation, would require 
$R_{e/p}\gtrsim 0.2$ for Coma and Abell 2256
and $R_{e/p}\gtrsim 0.025$ for Abell 2199.
Since measurements for galactic CRs 
give $R_{e/p}\sim 0.01-0.05$ \citep{muta87}, it would appear
that only the case of Abell 2199 could be explained in terms of
IC emission from primary electrons. However, as already pointed out,
our plot in the bottom right panel of Fig. \ref{HXRic.f} 
exhibits a large scatter, of about one order of
magnitude around the average values at each temperature, 
due to the different accretion/merger status of the simulated groups/cluster.
Thus the HXR emission from Coma cluster and Abell 2256 is also consistent with
the primary model, if a major accretion/merger process occurred in the
recent history of these clusters, as indeed indicated 
by the observations \citep{brieletal91,brlk94}.
 
As regarding the spatial distribution of the HXR radiation, it
is unknown for the Coma cluster \citep{fufeetal99},
but there is some indication that it may be more widely
distributed than the thermal emission
for Abell 2199 \citep{kaastraetal99}, as it would result 
from the IC process \citep{sali98}.
The emission from simulated secondary CRes is smoother and less extended than 
that from primary CRes, but the available information on the 
spatial distribution of detected HXR emission is not 
sufficient yet to discriminate observationally 
between the two alternative models.

\subsubsection{Extreme Ultra-violet Emission}

Regarding the EUV radiation, 
in Table \ref{euvmp.tab} we compare the
published EUV excess measurements in column 3 
with our numerical results for: the EUV 
luminosity due to IC emission from secondary CRes in column 4;
and the value of $R_{e/p}$ required to match the observations 
for primary CRes in column 5.

The values computed in Table \ref{euvmp.tab} for 
a cluster with the same temperature as Coma,
indicate that both primary and, perhaps even, secondary CRes could produce
through IC emission an EUV radiation flux comparable to the
measurement reported by \citet{bobeko99}.
Our numerical model, however, falls short of the detected emission 
from Virgo cluster \citep{beboko00,lieuetal96a} by about two orders of 
magnitude, although the presence of M81 may induce higher non-thermal 
particle populations in the ICM of Virgo and enhance the emission 
in this wave-band. For the remaining two clusters 
Abell 1795 and Abell 2199, our numerical model 
for both primary (assuming $R_{e/p}\sim 10^{-2}$) 
and secondary electrons, indicates a luminosity of order of 
a few $10^{41}-10^{42}$ erg s$^{-1}$,
well below the reported detections.
The two order of magnitude difference is significant even considering
the scatter in the reported plots and the 
limitations in our work. So for these two clusters we 
suspect that the reported EUV excesses
are unlikely to be produced by 
IC scattering off cosmic microwave background photons of
CRes shock-accelerated in the ICM.
This conclusion is further supported by the following insight.
The simulation results in Figs. \ref{euvic.f}
show a clear trend of the EUV luminosity increasing 
with the temperature of the collapsed object
for both primary and secondary electrons.
Such a trend is not present in the observational data. According 
to Table \ref{euvmp.tab},
the reported EUV luminosity for both Abell 1795
and Abell 2199 are larger by about two orders of magnitude than
the value reported for Coma by \citet{bobeko99}.
This contradicts the theoretical trends shown in our plots 
and cannot be accounted for by the scatter among the simulated groups/clusters.
We note that this peculiar feature of the observational 
results will affect any model in which the 
EUV radiation is produced by energetic electrons generated through
general cluster evolution. In fact, 
whether accelerated by accretion shock waves, merger shock waves, 
intra-cluster turbulence or generated in 
hadronic interactions, the energy content of the relativistic 
electrons can always be expressed as a
fraction of the total energy of the cluster 
(which scales with its temperature).

The synthetic EUV images can also be compared to the thermal X-ray and synchrotron
radio emission maps displayed on the same scale in Figs. \ref{radio2s.f}
and \ref{radio2e.f}.
As we can see, the EUV emissivity from both primary 
and secondary electrons is more extended than the thermal X-rays,
as observed in real clusters. However,
the EUV radiation in these IC based models is
also more extended than the radio emissivity; this 
contradicts observational results, as 
first pointed out by \citet{bobe98}. Thus,
the morphological properties of the emitting region pose 
another challenge to our attempt to attribute the EUV excess
to CRes of cosmological origin. 

\subsection{Synchrotron Radiation} \label{radio.se}

\begin{figure}
\plotone{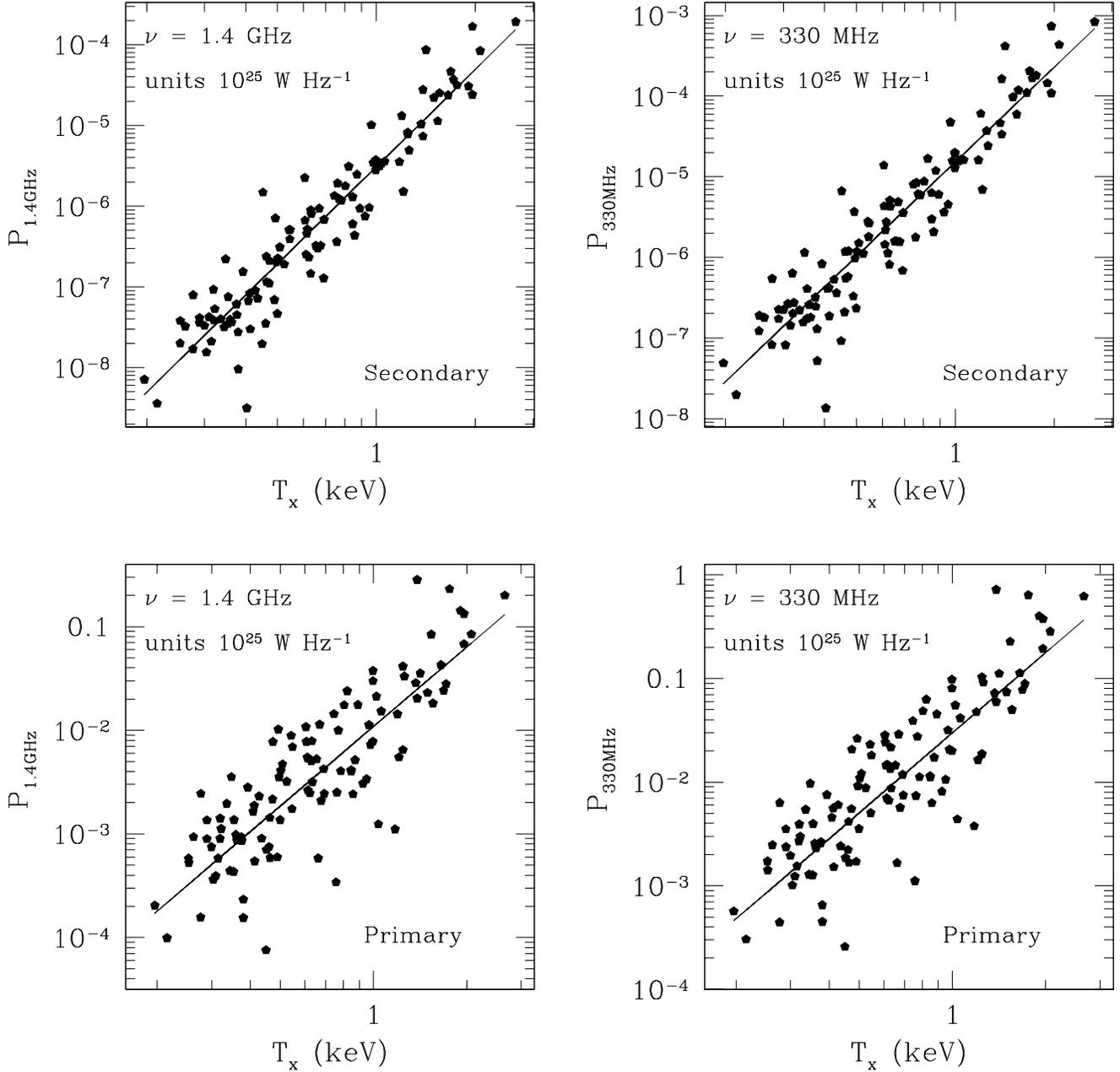}
\figcaption{Group/cluster synchrotron power 
at 1.4 GHz (top-left) and at 330MHz (top-right) from secondary electrons 
and at 1.4 GHz (bottom-left) and at 330MHz (bottom-right) from primary
electrons as a function of the core temperature. 
The emissivity is integrated over a spherical volume 
within a radius of 1.3 \hinv Mpc.
Note that the physical values of the luminosities for
primary electrons should be scaled with the electron to proton 
injection ratio, $R_{e/p}$.  
\label{syn.f}} 
\end{figure}

The synchrotron radiation emitted by CRes
has been calculated using eq. \ref{jsync.eq}. 
We remind the reader that we have assumed for this calculation a volume
averaged magnetic field in a Coma-like cluster of order 
3 $\mu$G and an emitting volume of radius 1.3 \hinv Mpc.
Fig. \ref{syn.f} displays 
the simulated radio spectral power, $P_{\nu}$, at 1.4 GHz (left) and 330 MHz (right)
from secondary (top) and primary (bottom) electrons, respectively,
as a function of the group/cluster temperature.
For each plot, the best fit parameters for a power-law 
correlation (\cf \S \ref{globp.se}) are reported in Table \ref{fit.tab}.
Although the emitting electrons
are approximately the same
as those producing HXRs,
the temperature dependence of $P_\nu$ is steeper now,
reflecting the additional scaling of the magnetic field
with the group/cluster temperature. 
The latter is close to $T_x^{0.75}$ (see \S 3.2.1) in our simulation,
and indeed the values of $\phi$ have increased by 
approximately one unit
with respect to the IC scattering emission.

\begin{figure}
\plotone{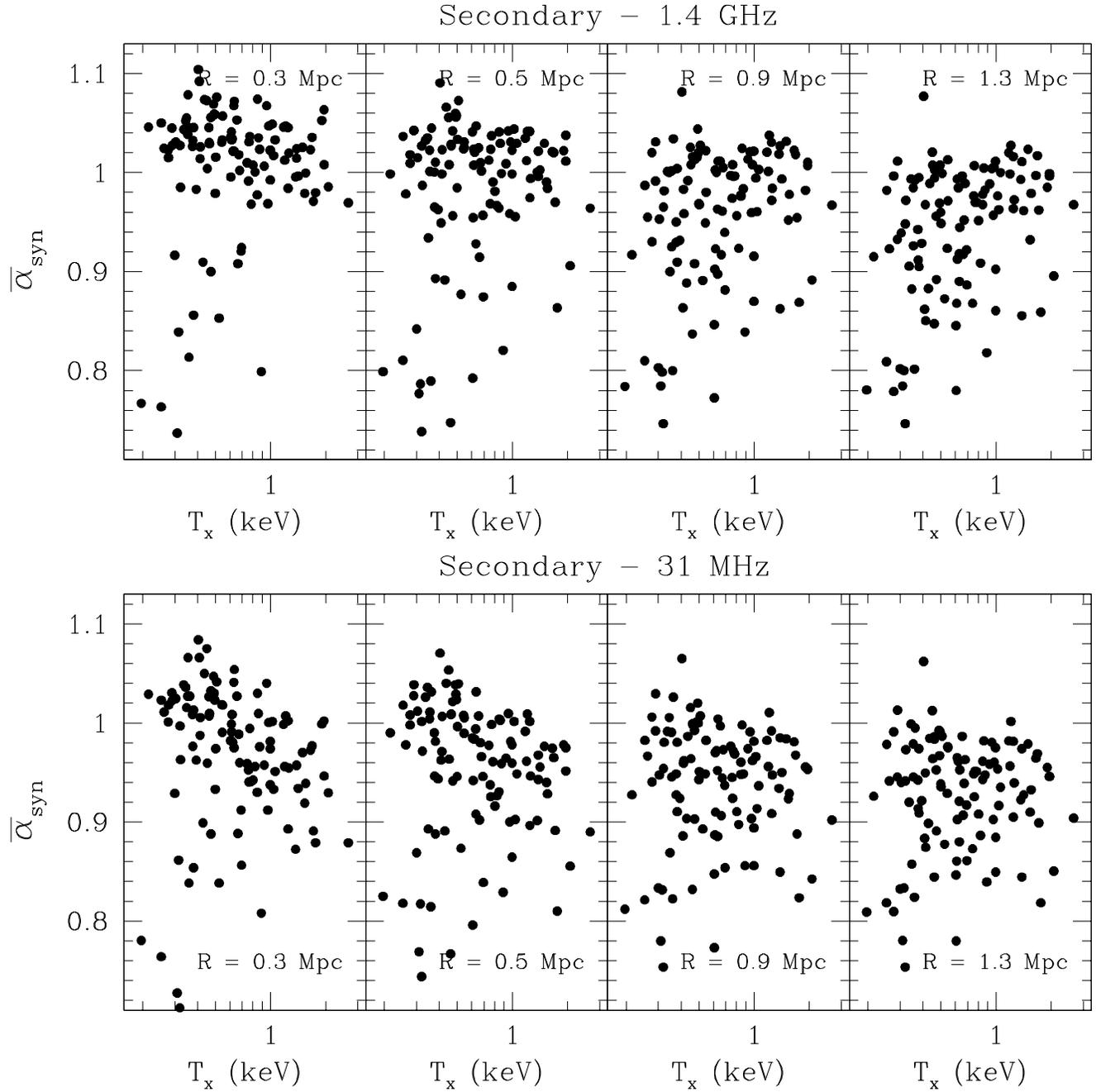}
\figcaption{Volume-averaged spectral index for the 
synchrotron radiation from secondary electrons 
at 1.4 GHz (top) and 31 MHz (bottom) 
as a function of the group/cluster core temperature. The four panels
in each row correspond to the following 
four different averaging volume radii: from left to right
$R= 0.3, 0.5, 0.9, 1.3$ \hinv Mpc.
\label{qsyns.f}} 
\end{figure}

\begin{figure}
\plotone{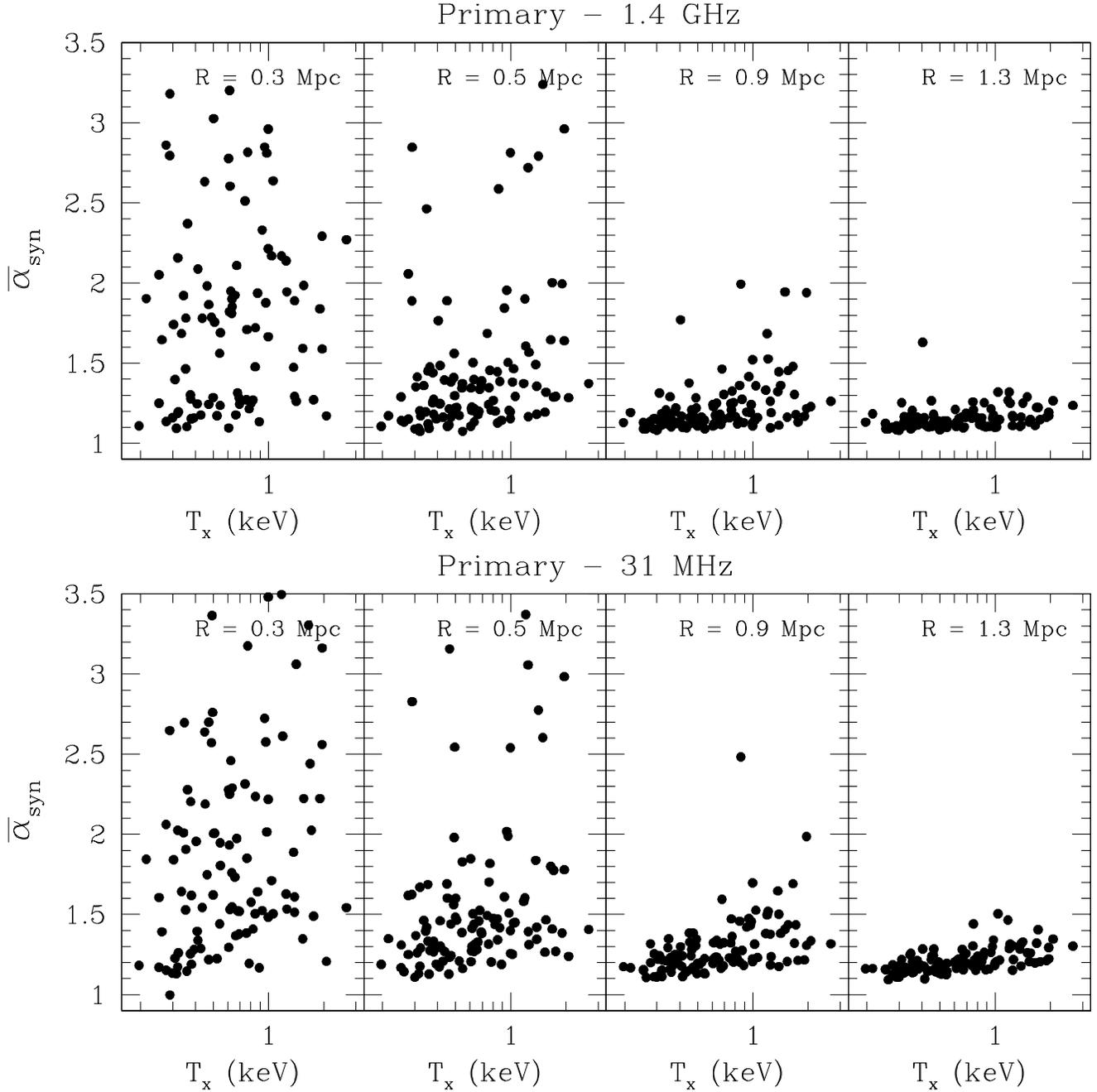}
\figcaption{As in Fig. \ref{qsyns.f} but now for
primary electrons. \label{qsyne.f}} 
\end{figure}

By analogy with the HXR emission, we have calculated the 
volume-averaged synchrotron spectral indices for radiation 
at 1.4 GHz and 31 MHz,
which bracket the range of most commonly observed frequencies.
These spectral indices are illustrated in Fig. \ref{qsyns.f}
for secondary electrons and in Fig. \ref{qsyne.f}
for the primary electrons. 
As for the HXR case, the spectral index associated
to the emission from primary electrons
represents the value averaged over the post-shock emitting volume.
As discussed at length in \S \ref{resic.se}, 
the spectral index of the CRes just behind the acceleration region 
should be smaller by one half with respect to what reported in those figures.
First of all, those plots 
do not show any relevant difference between values of the spectral index
at the frequencies of 1.4 GHz and 31 MHz, both in the
case of primary or secondary electrons.
This indicates that within the energy range that
correspond to the above frequencies, there is no
break due to radiative losses in the particle distribution.
Similarly to and with analogous considerations as for the HXR IC emission,
we find $ \bar{\alpha}_{syn}(<R) \gtrsim 1.1$ 
for the secondary electrons and $1-1.3$ for the primary electrons
in the large averaging volume which contains the acceleration sites.
Again, this is because the particles responsible for the emission 
in the two cases (radio and HXR) represent approximately
the same populations.

\subsubsection{Secondary Electrons and Radio Halos} \label{seelha.se}

The secondary electrons from our numerical simulation
can account for a number of features observed in radio halos
including the morphology of the emitting region, the polarization 
properties and the temperature dependence of the synchrotron power
at 1.4GHz \citep{lhba00}.

In Fig. \ref{radio2s.f} we show 
synthetic maps of the surface brightness from synchrotron
radiation from  
secondary electrons for the same object as in Fig. \ref{HXR2.f}.
In each of the three panels we present the radio emission
at different wavelengths, namely 1.4 GHz (top-left), 
330 MHz (top-right), 31 MHz (bottom-left). 
For reference, we also show the X-ray map produced by
the thermal bremsstrahlung. 
The radio emission extends throughout the X-ray group/cluster, 
indeed resembling the spatial features of observed radio halos.
By quantitative analysis of the emission map the lower 
frequency emission appears slightly 
more extended, again in accord with the observations.
This trend is due 
to the decay of the magnetic field strength at large distances 
from the cluster centers
and the fact that, according to eq. \ref{jsync2.eq},
the emissivity dims first at higher frequencies.
A comparison with the images of X-ray from thermal bremsstrahlung
reveals a higher level of
substructure in the radio emission than in the thermal emission.
Such sub-structure resembles 
to some extent that seen in the synthetic $\gamma$-ray images
of \citet{mrkj01} and is connected to the underlying shock distribution
in the ICM. 
However, the synchrotron map will also depend on the structure of the
magnetic field, which, in fully resolved simulations or in real clusters
might extend to regions smaller than we have been able to simulate
here. In any case, radio spatial features are also seen in real 
observations. There, the region responsible for the radio
emission often appears irregular and, in this sense, shows some 
departure from the observed X-ray image
(\eg Abell 665, \citealt{gife00}; Abell 520, \citealt{gefg00}; 
and Cl 0016+16, \citealt{gife00}).

According to our calculations, a typical Coma-like cluster ($T_x=8.3$ keV)
should exhibit a radio power at 1.4 GHz
of order $1.5\times10^{24}$ W Hz$^{-1}$, 
close to the observations (see Table \ref{radmp.tab}). 
We caution that this result depends on our choice of 
the magnetic field normalization of a few $\mu$G for a Coma-like cluster.
The significance of this finding is that the energy supplied by structure
shocks
would suffice to power radio halos if magnetic fields of a few $\mu$G
exist inside such clusters. In addition,
our numerical findings indicate a scaling of $P_{1.4GHz} \propto T^{4.2}$,
in good agreement 
with the observational values \citep{lhba00}. 
In fact, the 
synchrotron power in the secondary model should scale as
\begin{equation}
P_{syn} \propto N_{es} \, B^{1+\alpha_{syn}}  \propto
N_{cr}\,  n_{gas} \, \frac{B^{1+\alpha_{syn}}}{B^2/8\pi+u_{cmb}}
\propto T^{3} \, T^{2\phi_B} \propto T^{3+2\phi_B}
\end{equation}
where $\alpha_{syn} \simeq 1$ is the radio spectral index,
$N_{es}\propto N_{cr}\, n_{gas}/(B^2/8\pi+u_{cmb})$ is the number of secondary 
electrons, proportional to the product of the of cosmic 
ray ions and the gas density.
The factor $1/(B^2/8\pi+u_{cmb})$ 
simply accounts in a steady state for the reduction of secondary electrons from radiative losses.
Also, $\phi_B$ indicates parametrically
the temperature dependence of the magnetic field strength,
\ie $B\propto T^{\phi_B}$.
In \S \ref{hxr.se} we already saw that
$N_{es}\propto N_{cr}\, n_{gas}\propto T^3$.
In addition, from our simulation we have found $\phi_B \sim 0.75$.
The amplification of the magnetic field is primarily 
due to shear flows developed during large scale structure 
formation and, to a lesser extent, to gas compression.
The accuracy of the magnetic field strength dependence 
on the group/cluster temperature in the simulation 
is not clear, since
the $B-T_x $ relation is observationally unknown,
although now under investigation
\citetext{Kronberg, priv. comm.}.
Clearly, though, the coarse grid effects limit the amount of shear and
reduce the growth of the field, as it has already been pointed
out in analogous studies \citep{rostbu99,dobale99}. 

With the above temperature scalings of $N_{es}$ and $B$, the result 
$P_{1.4GHz} \propto T^{4.2}$ is roughly retrieved.
However, since $\alpha_{syn} \sim 1$,
when $B \ge (8\pi u_{cmb})^{1/2}(1+z)^2 \simeq 3.3 \mu$G,
the synchrotron power becomes independent of the magnetic field
strength and its dependence on the cluster temperature 
inevitably flattens. For the temperature scaling 
properties of the magnetic field in our simulation and the 
adopted normalization, such a turnover should occur at a
temperature above 9.4 keV (at $z=0$). Given the poor statistics of real clusters
with temperature beyond this value, conclusions in this respect
are premature.

We have also investigated the 
fractional polarization that characterizes the integrated
synchrotron emission in the synthetic maps. 
From our investigation we find that
the degree of polarization is very low throughout the simulated 
groups/clusters, typically of order of a few \%.
This result holds for the radiation at 1.4GHz and 330 MHz. 
In this calculation,
no Faraday depolarization was included. Its inclusion 
would further reduce the degree of polarization found here
\citep{doen00}, reinforcing our conclusions below. 
The low degree of polarization of synchrotron emission
is one of the distinctive features of observed radio halos.
The outcome in our model is probably due to the highly 
complex structure of the magnetic field inside the group/cluster.
In fact, the radiation from different depths along the line of sight 
will imprint a polarization angle determined by the integrated
structure of the field.
Since the field is a varying function of
space, the sum of all the contributions along one line of
sight and orthogonal field projections of equal magnitude
lead to a null net polarization, 
the emergent radiation from a disordered field will be mostly
unpolarized. 

Another important quantity that is determined by the observations is
the spectral index of the radiation. In the secondary CRe model 
this should be $\gtrsim 1.1$. For the case of the Coma cluster,
the best studied example, the integrated radio flux below
1.4 GHz is well fitted by a spectral index 1.16-1.3
\citep{deissetal97}.
This is marginally consistent with our results 
and, in fact, our current model underestimates 
the radio emission of the Coma cluster at 330 MHz
with respect to the data of \cite{vegife90}.
Additionally, the observed spectral index shows a steepening with 
increasing radius. No conclusion can be drawn from our plots in 
this respect. In fact, although 
Fig. \ref{qsyns.f} shows that the emission averaged 
spectral index tends to decrease slightly 
when larger volumes are considered, this is a spurious effect
as was discussed in \S \ref{resic.se}. 
In general the observational values of the 
spectral index may be slightly higher than what we predict 
\citep{hanish82}. In the context of
secondary electrons  from cosmic shocks this would correspond 
to slightly weaker shocks 
in the ICM than those typically found within group/cluster ICMs in
our simulation. In fact, there is a possible tendency for the Mach numbers 
we compute
to be slightly overestimated, especially in regions close to the cluster
core where with moderate spatial resolution the shock compression can 
be ``confused'' by the
gravitational compression \citep[see][for details]{minetal00}.
Although a precise estimate should be carried out through higher 
resolution simulations, a simple estimate indicates that 
\begin{equation}
\Delta\alpha = \frac{\Delta q}{2} = 
-1.5\; \frac{\gamma_{gas}+1}{(1-M^{-2})^{2}} \; \frac{1}{M^3} \; \Delta M,
\end{equation}
where $\gamma_{gas} = 5/3$ is the adiabatic index of the gas.
Therefore, an error $\Delta M\sim 1$
in the Mach number estimate translates into an error 
$\Delta \alpha \sim -0.2$ for the spectral index, which would 
allow for spectra steeper than reported in our plots.

\subsubsection{Primary Electrons}

In Fig. \ref{radio2e.f} 
we present synthetic maps of the surface 
brightness due to synchrotron emission
from primary electrons
at 1.4 GHz (top-left), 330 MHz (top-right) 
and 31 MHz (bottom-left) for the same object
of Figs. \ref{radio2s.f}.
On the bottom-right panel, we also present on the same scale 
a two-dimensional cut
of the contour levels of shock compression in a plane 
through the center of the collapsed object and perpendicular to
the line of sight of the other three panels.
In contrast to the emission produced by the secondary electrons, 
the primary electron emission maps are strikingly inhomogeneous,
with the primary electrons (and emission)
strongly concentrated around shocks.
Fig. \ref{radio2e.f} resembles
the HXR images presented in Fig. \ref{HXR2.f} (right panel), 
which also refers to the same object.

The location of these synchrotron emission regions at the immediate
outskirts of the group/cluster together with their irregular and
extended morphology, may suggest a connection with the
observed radio relics. The suggestion is reinforced by the 
finding that the fractional polarization through the 
extended regions is quite high, of order 30-50 \%,
comparable to the values suggested by the observations.
Unlike the the secondary electrons, the primary electrons
produce emission primarily in the immediate post-shock 
regions. There the magnetic field is highly aligned due to the 
enhancement of the perpendicular component by the shock compression.
Thus the radiation suffers only minor structural depolarization 
along the line of sight and emerges highly polarized. 
We do point out that our results might
overestimate the polarization expected in real clusters because finite
resolution limits the possible complexity of the magnetic field structure.

In Table \ref{radmpe.tab} we present a comparison 
between the observed properties of radio relics for a few
sources and the averaged values predicted in our numerical work.
To define the scope
of this comparison, we point out again that the predictions of our
model, illustrated in the bottom panels of Fig. \ref{syn.f}, 
indicate a range of values which span 
a factor of a few about some average. 
With this in mind, we notice that in all of the presented cases 
the mean predicted total power at 1.4 GHz is consistent with 
the observed values for a conservative estimate of $R_{e/p}$
of $10^{-2}-10^{-3}$.
This result signifies that, again,
the power supplied at shocks by accretion flows
is sufficient to generate the emission observed in extended
radio relics, for conservative values of shock 
acceleration injection and efficiency.

We have also calculated the volume-averaged spectral index  
(\cf \S \ref{resic.se}) for each pixel in the 
images of Figs. \ref{radio2e.f}. We find that in 
the brightest regions $\bar{\alpha}_{syn} \sim 1-1.3$ at both 1.4 GHz and
31 MHz. Where the emissivity is low,
the spectral index steepens to values 
$\bar{\alpha}_{syn} \approx 1.4-1.5$ (at 31 GHz).
These values are typical for all the groups/clusters analyzed with the 
simulation data \citep{mint00} and are
indeed consistent with the observed values for relics, 
a few of which are reported in table \ref{radmpe.tab}.
In addition, 
the above volume-averaged values of $\bar{\alpha}_{syn}$
corresponds to a spectral index of order $\sim 0.5-0.8$ 
for the bright emission regions
associated with the freshly accelerated populations
just behind the shock (\cf \ref{resic.se}).
Observation of the relic source 2006-56 is consistent with the above 
depiction as the spectral index increases from 
$\bar{\alpha}_{syn} \approx 0.5$ from the edge of the
source to about $\bar{\alpha}_{syn} \approx 1.1$ 
toward its center \citep{rottge97}.
This trend may indicate that the emitting 
particles in that source are indeed accelerated
by shocks, as was pointed out by \citet{ebkk98}. 

\section{Discussion and Conclusions} \label{disco.se}

In this paper we have explored the acceleration, 
losses and spatial transport of cosmic ray electrons
for the first time in a fully cosmological simulation of
large scale structure formation.
In our simulation, CRes were either 
directly injected at shocks (primary) or 
generated in hadronic collisions of CR ions and
the nuclei of the ICM (secondary).
In general the populations of CRs thus generated
are described by power laws, $f(p) \propto p^{-q}$,
with index $q$ depending on the strength of the shock. 
Our results indicate that the total number of relativistic electrons 
present in each group/cluster scales with the 
temperature of the ICM in the sense that 
more numerous populations are present in hotter groups/clusters.

Non-thermal emissions due to relativistic electrons
have been computed at different wavelengths in 
order to address the issues related to 
the recent EUV/HXR observations of clusters of galaxies.
As for the excess of emission in the EUV wave-band,
we find that the electrons generated in our simulation
are probably insufficient to produce the claimed 
detections \citep{lieuetal96a,lieuetal96b,mll98,beboko00},
except for the case of Coma cluster reported by \citet{bobeko99}.
Furthermore, the EUV excess reported so far has shown 
no trend with cluster temperature, contrary to the 
results of our simulation. 
Absence of such trends would suggest that processes 
responsible for generation of
non-thermal electrons do not depend on the energetics of groups/clusters.

The contribution to the HXR emissivity from the 
secondary electrons amounts to only a few 
percent of the currently reported values 
(namely, for the Coma cluster and Abell 2556 in the bandwidth 20-80 keV and 
Abell 2199 in the bandwidth 0.13-100 keV).
Increasing the total number of secondary electrons
enough to account for all of the HXR excess 
emission would imply a corresponding increase of parent 
CR ions and a consequent 
$\gamma$-ray flux much above the upper limits established by 
EGRET \citep{blco99,blasi99,mrkj01}. 

\begin{figure}
\plotone{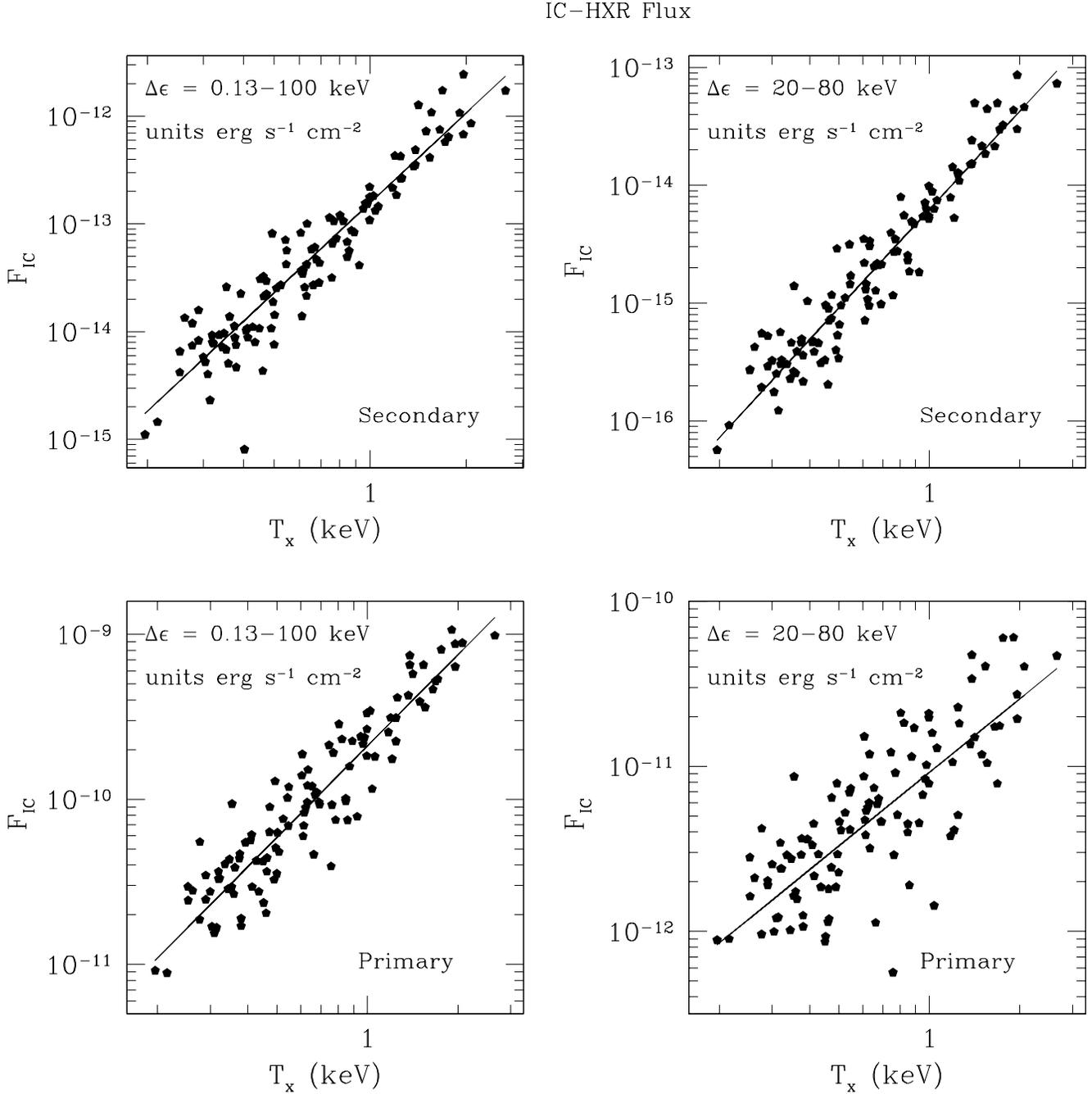}
\figcaption{Group/cluster HXR flux of IC emission 
in 0.13-100 keV (left) and in 20-80 keV (right) 
due to secondary (top) and primary electrons (bottom), respectively, 
as a function of the group/cluster core temperature. 
The emissivity is integrated over a spherical volume 
within a radius of 1.3 \hinv Mpc.
Note that the physical values of the luminosities for
primary electrons should be scaled with the electron to proton 
injection ratio, $R_{e/p}$.  
\label{HXRic.f}} 
\end{figure}

The average HXR emission from primary electron extrapolated 
from our plot in the bottom right panel of Fig. \ref{HXRic.f}
indicates a contribution from 
this component of about 10\% or so of the reported 
detections in this wave-band. This plot, however, 
is characterized by a large spread of about an order of
magnitude around the average, a reflection of the different 
level of shock activity in the ICM of individual groups/clusters.
Thus, the reported HXR emission from Coma cluster and Abell 2556
could be accounted for by IC emission of primary electrons,
given that these clusters are known to
be respectively in a post-merger phase \citep{brlk94} or in the 
process of merging \citep{brieletal91}.

From the analysis of the synchrotron emission, we find that our
secondary CRes could explain several features of radio halos; namely the 
total power at 1.4 GHz, the fractional polarization, and the
morphology and, marginally, even the spectral index.
The radio power, $P_{1.4GHz}$, as reported by \citet{lhba00},
for clusters at different temperatures
is produced by the secondary electron model,
provided the
magnetic field varies with the group/cluster temperature as steeply as 
found in our simulation. Consistency here is achieved
by assuming a normalization for the magnetic field 
that gives the volume-averaged value of 3 $\mu$G for a Coma-like cluster, 
as inferred from Faraday rotation measure estimates \citep{kkgv89,kkdl90}.
Numerous properties of radio halos are reproduced 
quite naturally according to our simulation,
although the spectral index reported by observations
is sometimes steeper than found here
\citep[see also][]{chietal96}. 
Future higher resolution simulations should allow us to determine 
whether or not this is a numerical artifact, as suggested in \S \ref{seelha.se}.
A similar result about the secondary model for radio halos
was recently obtained by 
\citet{doen00}, although in their case the 
CR population was not simulated but assumed from the
beginning. 
In our simulation the spectra of accelerated particles are determined
consistently with the cosmic shocks 
responsible directly or indirectly for their acceleration. 
Thus, particle distribution and spectra are not free parameters of the computation.

Finally, we have shown that the primary electrons directly accelerated at
cosmic shocks generate radio maps that are similar in radiated
power, morphology 
and polarization fraction to observed radio relics. 
The accelerating shocks have been shown to be strong enough 
to produce quite flat populations of CRes in the region
immediately adjacent to the acceleration sites. 
Our limited numerical resolution prevents
us from reproducing correctly the spatial distribution of
electron populations of different ``ages'' propagating away from 
the shock. 
However, the computed flat spectra of the shock-accelerated electron distributions 
indicate that the volume-averaged spectral index of the
whole post-shock emission should be in the observed range,
$\alpha =  1-1.5$.  

Thus the energetics of accretion/merger
shocks are in principle sufficient to power both the 
radio-relic emission and the reported detections of HXR 
excess. However, in order for the same electron population
to account for the emission in both wave-bands for either 
Coma cluster or Abell 2256 (see table \ref{hxrmp.tab} and \ref{radmpe.tab}),
it is required a volume-averaged magnetic field strength about an 
order of magnitude smaller than what assumed here.
It is conceivable that
the magnetic field strength drops by this amount as we move 
from the cluster center to its outskirts where the accretion 
shocks powering both the radio-relics and the HXR emission, 
would be located according to our simulations. 
And in fact this is supported by the magnetic field maps of 
our simulations \citep{mint00}. However, due to the limited numerical 
resolution such properties of the magnetic field cannot be 
considered as conclusive.

It is important to mention the existence of alternative models for
the existence of radio relics. In particular, it has been pointed out
that radio relics could be sites where fossil radio plasma, expelled by
radio galaxies at some point during cosmic history, is currently being 
revived through the encounter with a shock wave by means of adiabatic 
compression \citep{ensslin99,engo01}. The idea is compelling particularly 
because the ejection of relativistic plasma by radio galaxies 
in the intergalactic medium is an observational fact. It is shown that
radio plasma up to 2 Gyr of age can be revived by strong shocks \citep{engo01}.
The statistics, \ie the probability of actually seeing revived radio plasma at
a given epoch, is not an easy task because it
depends on the luminosity function of radio galaxies as well
as the subsequent evolution of the radio plasma in cosmic environment. 
However, future low frequency observation should improve substantially
our knowledge of the distribution of old radio plasma in cosmic environment \citep{engo01}.
Nevertheless, there exist some important differences between radio relics produced by primary 
electrons accelerated at shocks and those resulting from revived radio plasma.
In fact, according to
recent magetohydrodynamical simulations of the interaction of radio plasma
with a shock wave, 
a cloud of radio plasma becomes unstable upon being shocked,
and develops into a complex filamentary structure \citep{enbr01}. 
It is not clear that such 
morphological properties belong to giant radio relics observed at the outskirts
of cluster of galaxies. However, 
recent radio observations have
shown that a sample of radio relics with sizes up to a few hundred kpc
observed at very high resolution in a number of galaxy clusters
possess just these morphological features \citep{srmae01}.
It is possible then that both mechanisms, namely shock acceleration and 
compression of old radio plasma, are at work in the ICM producing a 
large variety radio sources with distinct properties, as outlined above. 
This possibility, just speculated here, should be investigated in the future.

From a broader perspective, it is very important to understand
the actual origins of radio halos and relics, so one can assess
the level of non-thermal activity in 
the ICM. If, as provided in the present model,
radio halos are produced by secondary electrons, they reveal the
existence of a related population of CR ions that have developed 
during the formation of the large scale structure 
\citep{voahbr96,bbp97,mint00,mrkj01}. 
Such populations may contain a significant fraction of the 
total energy density of the ICM
and could be, therefore, dynamically 
important \citep{mint00,mrkj01},
with many cosmological implications.

\acknowledgments

FM wishes to acknowledge support from 
a Doctoral Dissertation Fellowship 
at the University of Minnesota and from a fellowship
provided by the Research Training Network of the 
European Commission
for the Physics of the Inter-galactic Medium. 
The work of FM and TWJ has been supported by NASA grant NAG5-5055,
by NSF grants AST96-16964 and AST00-71167, and by the University of Minnesota
Supercomputing Institute.
DR and HK were supported in part by grant 1999-2-113-001-5 from the
interdisciplinary Research Program of the KOSEF.
We are grateful to I. L. Tregillis for his ray-tracer code
and to I. V. Moskalenko and  A. W. Strong for 
providing their GALPROP routines.
FM thanks Sebastian Heinz for useful comments on the manuscript and
acknowledges several insightful discussion with Torsten En{\ss}lin.
Finally, we thank the referee, P. Blasi, for several 
constructive comments to the manuscript.
\bibliographystyle{apj}
\bibliography{papers,books,proceed}
\clearpage
\begin{table}
\caption{Temperature scaling relations \label{fit.tab}}
\begin{center}
\begin{tabular}{lccccc}
\tableline\tableline 
 & \multicolumn{2}{c}{secondary e$^-$} & & \multicolumn{2}{c}{primary
\tablenotemark{a} e$^-$} \\ 
 Energy Band  &    $K$   & $\phi$ & & $K$ &  $\phi$ \\ 
 \\ 
\tableline  
   &  (erg s$^{-1}$ cm$^{-2}$) &  & &   (erg s$^{-1}$ cm$^{-2}$) & \\
$F_{ic}$ {\rm HXR:~20-80~keV}\hfill    & $2.8\times 10^{-13}$& 2.9& & $8.0\times 10^{-11}$ &1.5 \\ 
$F_{ic}$ {\rm HXR:~0.13-100~keV}\hfill & $7.0\times 10^{-12}$& 2.9& & $3.6\times 10^{-9} $ &1.9  \\
\tableline 
   &  (erg s$^{-1}$ ) &  & &   (erg s$^{-1}$ ) & \\
$L_{ic}$ {\rm EUV:~65-248~eV}\hfill & $2.2\times 10^{41}$& 2.9& & $3.8\times 10^{43} $ &1.9  \\
\tableline 
   &  (W Hz$^{-1}$ ) &  & &   (W Hz$^{-1}$ ) & \\
$P_{1.4 {\rm GHz}}$\hfill & $6.3\times 10^{23}$& 4.2& & $7.3\times 10^{25} $ &2.6  \\
$P_{330 {\rm MHz}}$\hfill & $2.4\times 10^{24}$& 4.1& & $2.1\times 10^{26} $ &2.6  \\
$P_{74 {\rm MHz}}$\hfill & $1.2\times 10^{25}$& 4.2& & $7.4\times 10^{26} $ &2.7  \\
$P_{31 {\rm MHz}}$\hfill & $2.9\times 10^{25}$& 4.2& & $1.6\times 10^{27} $ &2.8  \\
\tableline 
\end{tabular} 
\tablenotetext{a}{ The values reported in the table for primary electrons 
assume $R_{e/p}=1$.} 
\end{center} 
\end{table}
%
%
%
%
\begin{table}
\tabletypesize{\scriptsize}
\caption{Flux of HXR radiation:
measurements vs predictions.\label{hxrmp.tab}}
\begin{center}
\begin{tabular}[b]{lccccccccc} \tableline \tableline
                &       &  &      &     &\multicolumn{4}{c}{predictions} &  \\
                &       \multicolumn{4}{c}{measurements}  &  \multicolumn{2}{c}{secondaries}   &\multicolumn{2}{c}{primaries\tablenotemark{a}} &  \\
 Cluster      & z & $T_x\tablenotemark{b} $ &
                $F_{HXR}\tablenotemark{c}$ & 
$\alpha\tablenotemark{f}$ & $F_{HXR}\tablenotemark{c}$
                &$\alpha_{ic}\tablenotemark{f}$  & 
$R_{e/p}\tablenotemark{g}$ & $\alpha_{ic}\tablenotemark{h}$  & reference(s)    \\  \tableline
Coma      \tablenotemark{d}\hfill & 0.0232 & 8.3 & 2.2 & 0.7-3.6 &$5.2\times 10^{-2}$  & $\sim$1.05& 0.2 & 1-1.3& \cite{fufeetal99} \\
Abell 2256\tablenotemark{d}\hfill & 0.0581 & 6.95& 1.2 & 0.3-1.7&$4.8\times 10^{-3}$& $\sim$1.05& 0.8  & 1-1.3&  \cite{fufeetal00} \\ 
Abell 2199\tablenotemark{e}\hfill & 0.0305 & 4.6 & 3.2 & 1.5-2.0& 0.13 & $\sim$1.05& 0.03 & 1-1.3&  \cite{kaastraetal99} \\
 \tableline
\end{tabular}
\tablenotetext{a}{We report here the required value for $R_{e/p}$ in
order to explain the observed emissivity with the primary electrons model.} 
\tablenotetext{b}{Temperatures are measured in keV.}
\tablenotetext{c}{Fluxes are measured in units $10^{-11}$ erg s$^{-1}$ cm$^{-2}$.}
\tablenotetext{d}{For Coma cluster and Abell 2256 all fluxes correspond to the bandwidth 20-80 keV.}
\tablenotetext{e}{For Abell 2199 cluster all fluxes correspond to the bandwidth 0.13-100 keV.}
\tablenotetext{f}{The spectral index $\alpha_{ic}$ is defined in \S  \ref{spidx.eq}.}
\tablenotetext{g}{We report here the required value for $R_{e/p}$ in
order to explain the observed emissivity with the primary electrons model.}
\tablenotetext{h}{The reported values of the spectral index $\alpha_{ic}$ refer to
averages over the post-shock emitting volume (\cf \S \ref{resic.se}). 
For freshly accelerated
particles just behind the shock they should be in the range $0.5-0.8$.}
\end{center}
\end{table}
%
%
%
%
%
\begin{deluxetable}{lccccc}
\tablecaption{Emission of EUV radiation in the EUV band~65 - 248 ~eV:
measurements vs predictions.\label{euvmp.tab} }
\tablewidth{0pt}
\tablehead{  
 & & & 
\multicolumn{2}{c}{Predictions\tablenotemark{a}}
& \\
\colhead{Cluster} &
\colhead{ $T_x$} &
\colhead{measurements} &
\colhead{secondaries} &
\colhead{} &
\colhead{reference(s)} 
 \\
 &
\colhead{keV}  &
\colhead{(erg s$^{-1}$)}  &
\colhead{(erg s$^{-1}$)}  &
\colhead{$R_{e/p}$}  &
}
\startdata
     
Coma (A 1656)\hfill & 8.3  & $1.5 \times 10^{42}$ & $4.0\times 10^{41}$& $6\times 10^{-3}$& \cite{bobeko99} \\
             \hfill & 8.3 & $5.0 \times 10^{43}$ & $4.0\times 10^{41}$& $0.2$ & \cite{lieuetal96a} \\
Virgo \hfill        & 1.8 & $5.2 \times 10^{42}$ & $4.8\times 10^{39}$ & $0.4$ & \cite{beboko00}\\
      \hfill        & 1.8 & $9.0 \times 10^{42}$ & $4.8\times 10^{39}$ & $0.7$ & \cite{lieuetal96a}\\ 
Abell 1795\hfill    & 5.8 &   n. d.              & $1.4\times 10^{41} $ & $-$ & \cite{bobeko99}\\
      \hfill        & 5.8 & $2.0  \times 10^{45}$ & $1.4\times 10^{41} $ & $18$ & \cite{mll98}\\
Abell 2199\hfill    & 4.5 &    n. d.             & $7.0\times 10^{40} $ & $-$ & \cite{bobeko99} \\
    \hfill          & 4.5 & $5.0 \times 10^{43} $ & $7.0\times 10^{40} $ & $0.7$ & \cite{sake00}
\enddata

\tablenotetext{a}{We report here the required value for $R_{e/p}$ in
order to explain the observed emissivity with the primary electrons model.}
\end{deluxetable}
%
%
%
%
\begin{table}
\caption{Radio power at 1.4 GHz:
measurements vs predictions of secondary electrons model.\label{radmp.tab}}
\begin{center}
\begin{tabular}{lccc}  \tableline \tableline 
  Quantity  & measurements      & model & reference(s)  \\ \tableline 
$P_{1.4GHz}$(8.3keV)\tablenotemark{a}& 1.5               &  1.5  &   \cite{deissetal97}  \\
$\phi$    & 4-5.2             & $\sim 4.2$  &\cite{lhba00}   \\   
$\alpha_{syn} \tablenotemark{b}$       & 0.5-$\le$ 1.6     & $\gtrsim 1.1$ &  \cite{hanish82,deissetal97} \\ 
Pol.        & $\le $ 10 \%      &  a few \%  &\cite{feretti99}   \\ 
\tableline 
\end{tabular} 
\tablenotetext{a}{ The radio power is measured in units $10^{24}$ W Hz$^{-1}$.
Here we compare the normalizations of the $P_{1.4GHz}$ vs $T_x$
relation by taking the measured and predicted values of $P_{1.4GHz}$
for the Coma cluster.}
\tablenotetext{b}{The range includes values for different clusters as well as 
spatial variations inside the same cluster.} 
\end{center}
\end{table}
%
%
%
%
\begin{table}
\caption{Properties of Radio Relics:
measurements vs predictions of primary electrons model.\label{radmpe.tab}}
\begin{center}
\begin{tabular}[b]{lcccccccc}\tableline\tableline  
   &   &  & \multicolumn{3}{c}{measurement} && \multicolumn{1}{c}{model\tablenotemark{b}} &   \\ 
\cline{4-6}\cline{8-8} \\ 
   Relic & Cluster & freq.  & Flux\tablenotemark{a} & Pol.     &   $\alpha$     & &$R_{e/p}\tablenotemark{c}$ &  ref.(s)  \\ \tableline
1253-275 & Coma    & 1.4 GHz & 0.16     & 25-30 \% & $1.1\pm 0.2\tablenotemark{d}$ & &$1\times 10^{-3}$ &  \cite{vegife90,gifest91}  \\
         &         & 330 MHz & 1.4      & n. a.    & $1.1  \pm 0.2$ & & $2\times 10^{-3}$  &  \cite{gifest91}  \\
2006-56 & A 3667 & 1.4 GHz & 0.7    & n. a.    & $0.5-1.2 $ && $ 4\times 10^{-2} $  & \cite{gossetal82,rottge97}\\
Comp. G & A 2256 & 1.4 GHz & 0.15  & 20 \%     & $0.8\pm 0.1$ & &$ 6\times 10^{-3} $ &   \cite{rottge94}  \\
Comp. H & A 2256 & 1.4 GHz & 0.1   & 20 \%     & $0.8\pm 0.1$ && $5\times 10^{-3}$ &  \cite{rottge94}  \\
1140+203 & A 1367 & 1.4 GHz & 0.18   & n. a.   & $1.9\pm 0.5$ && $5\times 10^{-3}$ & \cite{gatr83}  \\
%
%
\end{tabular} 
\tablenotetext{a}{The radio flux is measured in Jy = 10$^{-23}$ erg cm$^{-2}$.}
\tablenotetext{b}{The spectral index in the model is always
in the range 0.5-0.8 for the emitting region immediately behind the shock and
1-1.3 when the volume-average value is considered (see \S \ref{resic.se} for 
further details). Likewise, the prediction for the fractional
polarization is always 30-50 \%, although geometrical effects may reduce 
the observed value with respect to it \citep{ebkk98}.}
\tablenotetext{c}{Here we report the required value for $R_{e/p}$ in
order to explain the observed emissivity with the primary electrons model.}
\tablenotetext{d}{\cite{anfegi84} find 1.18.} 
\end{center}
\end{table}
\clearpage
%
%


\begin{figure}
\figcaption{Emission map of HXR 
in units $2\times 10^{-4}$ erg cm$^{-2}$ s$^{-1}$ per pixel,
from secondary (left) and primary (right) electrons
at 50 keV
for one of the collapsed objects generated in the simulation.
The linear size of the panels correspond to 5.5 \hinv Mpc.
\label{HXR2.f}} 
\end{figure}

\clearpage 

\begin{figure}
\figcaption{Emission map in units $4\times 10{-7}$
erg cm$^{-2}$ s$^{-1}$ per pixel,
of EUV radiation 
for the same object as in Fig. \ref{HXR2.f}.
Left and right panels correspond to secondary and primary 
electrons, respectively.
The linear size of the panels correspond to 5.5 \hinv Mpc.
\label{euv2d1.f}}
\end{figure}

\clearpage 

\clearpage 

\clearpage 

\clearpage 

\clearpage 

\clearpage 

\clearpage 

\begin{figure}
\figcaption{Radio map of surface brightness in units
of Jansky per pixel (1 Jy = 10$^{-23}$ erg cm$^{-2}$)
for the same object as in Fig. \ref{HXR2.f},
at 1.4 GHz (top-left), 330 MHz (top-right)
and 31 MHz (bottom-left). The radiation is synchrotron emission
generated by secondary electrons in the group/cluster magnetic field. 
For comparison we also show the thermal X-ray map (bottom-right)
in units erg cm$^{-2}$ s$^{-1}$. 
The linear size of the panels correspond to 5.5 \hinv Mpc.
\label{radio2s.f}} 
\end{figure}

\clearpage 

\begin{figure}
\figcaption{Same as for Fig. \ref{radio2s.f} but for
primary electrons in the top-left, top-right and bottom left panels. 
A two-dimensional cut of the contour levels of shock compression 
in a plane through the center of the collapsed object is also shown on
the bottom-right panel for comparison. \label{radio2e.f}}
\end{figure}

\end{document}